# Participation and Division of Labor in User-Driven Algorithm Audits: How Do Everyday Users Work together to Surface Algorithmic Harms?

Rena Li[*]
Sara Kingsley[*]
renal@andrew.cmu.edu
skingsle@cs.cmu.edu
Carnegie Mellon University
Pittsburgh, Pennsylvania, USA

Chelsea Fan[†]
Proteeti Sinha[†]
Nora Wai[†]
Carnegie Mellon University
Pittsburgh, Pennsylvania, USA

Jaimie Lee[‡]
jaimiele@andrew.cmu.edu
Carnegie Mellon University
Pittsburgh, USA

Hong Shen
hongs@andrew.cmu.edu
Carnegie Mellon University
Pittsburgh, USA

Motahhare Eslami
meslami@andrew.cmu.edu
Carnegie Mellon University
Pittsburgh, USA

Jason Hong
jasonh@cs.cmu.edu
Carnegie Mellon University
Pittsburgh, USA

## ABSTRACT
Recent years have witnessed an interesting phenomenon in which users come together to interrogate potentially harmful algorithmic behaviors they encounter in their everyday lives. Researchers have started to develop theoretical and empirical understandings of these user-driven audits, with a hope to harness the power of users in detecting harmful machine behaviors. However, little is known about users' participation and their division of labor in these audits, which are essential to support these collective efforts in the future. Through collecting and analyzing 17,984 tweets from four recent cases of user-driven audits, we shed light on patterns of users' participation and engagement, especially with the top contributors in each case. We also identified the various roles users' generated content played in these audits, including hypothesizing, data collection, amplification, contextualization, and escalation. We discuss implications for designing tools to support user-driven audits and users who labor to raise awareness of algorithm bias.

## 1 INTRODUCTION

The presence of biases in algorithmic systems has given rise to auditing approaches, usually led by AI/ML experts, to investigate these systems for harmful behaviors [17, 52]. These expert-driven auditing techniques have been successful in finding and mitigating many cases of harmful algorithmic behavior; yet, they suffer from a number of limitations. For one, experts are not always aware of emergent biases affecting marginalized communities in new ways. Further, expert techniques to investigate bias were not designed to necessarily detect the unpredictable behavior of algorithmic systems before or after a system is deployed. Instead, most expert-led auditing methods were developed to detect statistical disparities — not, for example, if an algorithm is censuring or harmfully depicting a marginalized community in images or the provision of online services.

Recent years have witnessed an interesting phenomenon of *user-driven audits* that can overcome some of these limitations. In user-driven audits, end-users organically come together and conduct audits of algorithmic systems to uncover, interrogate, and make sense of potentially harmful machine behaviors they encounter in their everyday lives [44, 63, 65]. Recent examples of user-driven audits include Twitter users detecting racial bias in its image cropping algorithm [41], small business owners coming together to investigate Yelp's potential bias against businesses that do not advertise with the platform [27], content creators evaluating why Youtube's algorithm demonetizes LGBTQ content published to the platform [44], and users testing Google Translate for gender bias [56]. Some of these user-driven audits have led major technology companies to change or stop using their algorithms in production [10, 58], or motivated lawmakers to propose new public policies, showing these investigations can have broad social impact [75].

---

[*]Both authors contributed equally to this research.
[†]Authors contributed equally to this research.
[‡]Author contributed the data visualizations and significant data analysis work.

Inspired by these developments, researchers have started to develop both theoretical and empirical understandings of user-driven auditing, with a hope to harness the power of end-users in detecting harmful algorithmic behaviors (e.g., [27, 28]). In particular, Shen and DeVos et al. [65] used exploratory case study to theorize how users detect, understand, and interrogate problematic machine behaviors via their daily interactions with algorithmic systems, as well as how user-driven audits can sometimes be more effective compared to audits led by experts. In another qualitative study, DeVos et. al. (2022) found during user-driven audits, users engage in sense-making, e.g. extrapolating theories from their observations about how an algorithmic system to explain the bias they perceived and proposed patterns of such sense-making process [23].

Despite these recent developments, however, little is known about the participation patterns and the division of labor among users in these algorithm audits, especially among a relatively large number of users. Analyzing participation patterns and division of labor in user-driven audits is essential if we want to help organize and support users in conducting these kinds of algorithmic audits in the future. Therefore, in our study, we asked the following three research questions: **(RQ1)** *what are the patterns of participation among users in user-driven algorithm audits*, **(RQ2)** *who are the top contributors in these audits* in terms of number of contributions and spreading the audit amongst their followers, and **(RQ3)** *what are the different roles that tweets play in these audits?*

To answer our research questions, we present a comparative analysis of four different user-driven audits on Twitter. We chose Twitter as our primary research platform because it allows us to access rich user-generated data, which enables easy comparison across different cases. In addition, Twitter has also been used as one of the major social media platforms for user auditors to share concerns around problematic machine behaviors in the past. For example, in the YouTube LGBTQ demonetization case [44], users eventually chose to publicly tweet about the machine biases after failing to receive a response from the system developer. We selected four high-profile cases, each with a relatively large number of participants, and spanning a spectrum of different auditing processes and origins to enable effective comparison. Those cases include the aforementioned Twitter's image cropping algorithm which seemed to exhibit racial bias in image previews [74]; the ImageNet Roulette art project which let people see the results of a computer vision classifier on uploaded images, often showing offensive labels [19]; Portrait AI's smartphone app that redraws photos using different famous painting styles, often redrawing people of different races as Caucasian; and Apple Card, a credit card that for married couples often gave men much higher credit limits than women [47]. In total, we collected 17,984 relevant tweets, qualitatively coded and analyzed 1,800 of them. We then identified and analyzed the top contributors in each case and developed five categories describing users' division of labor. Finally, we developed a machine learning classifier to label the remaining tweets and analyzed how frequently and over what time period users engaged in auditing activities.

Our analysis shed light on the patterns of users' participation and engagement across all the four cases. We also identified five major roles users' generated content played in these audits on Twitter, including hypothesizing (tweets that hypothesize why the observed machine behavior is happening), evidence collection (tweets that use evidence to support or oppose a hypothesis proposed by others), amplification (tweets that help share and broadcast information necessary for others), contextualization (tweets that place the ongoing audit within the broader social, technical or cultural context), and escalation (emotional tweets that primarily react to what's been observed within the audit). Our results have contributed and enriched our understanding of user-driven audits offered in the past literature [23, 49, 65] by showing that (1) instead of focusing on engaging deep and prolonged participation of users in audits [49], sometimes simple and short tweets can also play a vital role in demonstrating outrage and raising awareness, which might lead to system-level changes from algorithm operators and other authorities (e.g., government agencies); and (2) instead of focusing on scientific methods of hypothesizing and evidence collection [49], "amplification" and "escalation" seem to be especially critical for user-driven audits to engender broader awareness and to elicit a response from algorithm operators and government authorities.

In sum, our research offers three major contributions:

- *Comparative analysis of User Participation Patterns and Engagement:* We offer descriptive statistics of general patterns of user participation across all four cases, including the number of users participating, the length of time users engaged in audit related activities, as well as top contributors in each case.
- *Identification of the Division of Labor in User-generated Content:* We identify five general roles users' tweets played across all of the four user-driven audits, describing subdivisions of labor within each role as well as variations in terms of how these roles differ across the cases.
- *Design Implications*: Based on these empirical findings, we present a set of design suggestions for better supporting user-driven audits in the future, as well as a discussion of what it means for user-driven audits to be successful as compared to conventional types of algorithm audits.

## 2 RELATED WORK

In this section, we outline relevant work in two areas. First, we survey existing approaches to algorithms auditing and describe how our work is positioned in this space. Next, we review existing work on collective intelligence and sensemaking, and describe how our work on user-driven auditing draws from and contributes to this line of research.

### 2.1 Algorithm Audit

Over the past few years, a wide range of machine-learning algorithms have been criticized for biases and harmful behaviors. In response, researchers in HCI, CSCW and Machine Learning have developed a variety of approaches to inspect and investigate algorithmic systems. These approaches are often called "algorithm audit" [17, 52], referring to methods of "repeatedly querying an algorithm, or a software application relying on an algorithmic system, and observing its output in order to draw conclusions about the algorithm's opaque inner workings and possible external impact" [52].

Past research has successfully identified harmful and discriminatory behaviors in a variety of algorithmic domains, including search engines [52, 54, 60], online advertising [45, 67], facial recognition [10], word embedding [7], and e-commerce [40]. For example, Buolamwini and Gebru audited three commercial gender classification systems and found that the commercial systems misclassified darker-skinned women more often than white people [10]. In another study, Sweeney [67] audited Google's search and advertising system, finding that more ads for arrest records were displayed when names associated with Black people were searched for versus White people. These studies, however, were mostly conducted and centrally organized by expert researchers, with relatively high levels of technical expertise.

In recent years, another form of algorithm auditing has emerged, where regular users gather together to detect, hypothesize, and test for instances of bias in the systems they use [27, 28, 65]. Through a series of exploratory case studies, Shen and DeVos et al. [65] developed the concept of "everyday algorithm auditing" to conceptualize a process in which users interrogate problematic machine behaviors via their daily interactions with algorithmic systems. DeVos et al. [23] conducted a series of think-aloud interviews, diary studies and workshops, to more closely look at how users find and make sense of harmful machine behaviors. Their findings suggest users' auditing strategies are significantly guided by their personal experiences with and exposures to societal biases. These user-driven audits have also inspired experts to more rigorously test the underlying algorithms embedded into software applications. For example, inspired by users' efforts in detecting the racial and gender biases of Twitter's cropping algorithms, Yee et al. [74] provided quantitative fairness analysis showing the level of disparate impact of Twitter's model on racial and gender subgroups. Experts have also built tools to support user-driven audits. For example, Lam et al. (2022) [49] presented a system for "lowering the high effort threshold of algorithm auditing" and "scaffold[ing] the auditing process" for end-users.

The above shows that expert vs. user-driven is one dimension for characterizing algorithm audits. Costanza-Chock et al [17] introduce another dimension, differentiating between first-, second-, and third-party algorithm audits. First-party audits are conducted by the organization or developers that created the algorithm, with an explicit goal of systematically evaluating the algorithm for potential biases and harmful behaviors [17]. In second-party audits, paid experts who have access to the algorithm and/or backend data conduct the audit, reporting their findings back to the first party, and sometimes also the public. In contrast, third-party algorithm audits are initiated by independent entities that have no contractual relationship with the organization, e.g. independent researchers, journalists, or government agencies.

Combining these two dimensions, we can see that much of the past work in this space has been expert-driven and first-, second-, and third-party audits. In this paper, user-driven audits can be considered a new kind of third-party algorithm audit. However, one major difference is that these user-driven third-party audits are conducted in a form that is far less structured than expert-driven audits. Another sharp difference is that the many users involved in a third-party user-driven audit might not have an explicit or even a consistent goal. Despite this, at least three of the four cases we examine successfully led to some kind of change or intervention. We also note that it is possible to have user-driven first-party audits, e.g. where members of a company might try out a system internally [17, 65], as well as user-driven second-party audits, where a company might hire a crowd of workers to systematically test a system (see [11] for one example of a tool to help in this space). However, these two kinds of audits are outside of the scope of this current paper.

Despite these recent developments, there is currently little empirical analysis of many aspects of the real-world dynamics of user-driven audits. Past work [65] suggests that there are different types of user-driven audits, spanning across different levels of algorithmic expertise, collectiveness, and organicness. However, we currently have a limited understanding of how users participate in different types of algorithm audits. For example, what are typical participation patterns? What kinds of user-generated content do users contribute? Understanding these kinds of questions is helpful for future analyses of user-driven audits, and also necessary for building tools to support these kinds of user-driven audits. Drawing on mixed methods, this paper adds to this emerging line of research by empirically comparing four different auditing cases performed collectively by users on Twitter. Our results show that although these cases differ in a variety of ways, there are also many similarities in terms of contributors as well as categories of contributions.

## 2.2 Collective Intelligence and Sensemaking

Defined by Weick [71] as "placement of items into frameworks, comprehending, redressing surprise, constructing meaning, interacting in pursuit of mutual understanding, and patterning," the notion of sensemaking has been widely used to examine how a group of people can work together, build on each other's discoveries and insights, and work towards a common goal — sometimes yielding better results than domain experts working independently. Past work has explored collective intelligence and sensemaking across a range of contexts, including image labelling [70], knowledge mapping and curation (e.g., on Wikipedia) [30, 33, 38, 39], scientific research and collaboration [16, 18], and social commerce [13]. Another line of literature related to sense-making explores how users develop 'folk theories' to explain their experiences and observations of using computer and algorithmic systems [21, 37, 43]. Past work has investigated how these folk theories and existing perceptions can shape and influence people's attitudes and behaviors towards algorithmic systems [6, 9, 22, 25, 36, 68, 72]. A growing number of these studies focus on user theories of algorithm bias [22, 43, 44]. Recent work on user-driven algorithm audit has also discussed the role of folk theories in users' hypothesizing and testing the potential biases in algorithmic systems [65]. However, in this sensemaking process, little is known about what roles user take when evaluating and detecting potential algorithmic biases.

Our work aims to complement this line of work by looking into users' participation and division of labor across four different algorithm auditing cases on Twitter. In particular, we are interested in understanding how Twitter users collaboratively hypothesize, test, make sense of problematic algorithmic behaviors on social media platforms, and raise awareness of issues of harmful biases.

## 3 BACKGROUND

Here, we offer background information on the four auditing cases on Twitter that we examined, namely Twitter Image Cropping, ImageNet Roulette, Portrait AI, and Apple Card. We chose these four cases for a number of reasons [22, 57]. We chose Twitter as our primary research platform because it allows us to access rich user-generated data and enables easy comparison across different cases. We selected cases with relatively high visibility and with a non-trivial number of people commenting on issues of algorithmic bias, which gave us a relatively large amount of user-generated data to analyze. We also looked for cases that spanned a spectrum of different origins and auditing processes. For instance, the Twitter Image Cropping, Portrait AI, and Apple Card cases are primarily bottom-up, where it was not originally clear that the algorithm was biased, and users organically came together to test for biases. In contrast, ImageNet Roulette is more top-down, where the creators of the site designed the tool to expose the underlying biases in their algorithm. Furthermore, the Twitter Image Cropping case was relatively short-lived focusing on a specific issue, whereas Portrait AI is long-lived without a specific focus. Additionally, Twitter Image Cropping, ImageNet Roulette, and Portrait AI focus primarily on images, whereas Apple Card focuses on user experience and testimonies instead.

Following previous research in this domain [23, 65], we consider those cases as user-driven algorithm audits as they were conducted first and foremost by *users* of an algorithmic system. It is possible that user-driven audits will have varying degrees of algorithmic expertise, collectiveness and organicity; however, they differ from expert-led audits because the reported harmful machine behaviors could only be detected within authentic, situated contexts of AI usage [65]. We also follow the same lifetime proposed by [65], considering the range of activities that a user-driven audit encompasses would involve – on a high level – the following phases: initiation, awareness raising, hypothesizing & testing and remediation. In practice, however, a user-driven audit may follow a non-linear path, some may skip certain phases and some may end before reaching all phases, as will be shown in our analysis of the four cases.

### 3.1 Twitter Image Cropping Case

To keep the size of photograph thumbnails posted on the site uniform, Twitter previously used an image cropping algorithm that would automatically crop users' photos. On September 18 2020, PhD student Colin Madland found that an image he posted of himself and a Black colleague seemed to always favor his [Colin's] face (even when flipping the orientation and order of images), suggesting that the algorithm favored lighter skinned faces.[1] This Twitter thread sparked uproar, leading numerous users to conduct their own tests and find other examples of bias in Twitter's image cropping algorithm. One example test was to upload a single picture containing both President Barack Obama (dark skin) and Senator Mitch McConnell (light skin) in various layouts, leaving a large amount of whitespace between the two faces to force the algorithm to choose a single face. Some images had Obama on top and McConnell below, others vice versa. On cropping, the algorithm appeared to favor McConnell regardless of configuration. This specific technique of having only two faces widely separated in the same image was quickly adopted by other auditors [41].

In a blog post on October 1, 2020 [3], Twitter disclosed that the algorithm used was a saliency algorithm, meaning that the algorithm honed in on parts of a picture that people's eyes were most likely to be drawn to [3]. Twitter acknowledged that this algorithm could lead to harm including unequal treatment based on demographic differences, objectification biases, and a lack of freedom to express oneself on Twitter without an algorithm making the decision for them. Twitter issued a public apology in response to the public outcry about the biased image cropping algorithm.

Overall, this audit received a great deal of popular press (e.g., [2, 41, 42]). Largely due to the publicity of the results of this audit, in May 2021, Twitter completely changed their image cropping feature, removing the algorithm altogether and allowing users to upload pictures of any standard aspect ratio.[2] The tweet preview also shows the user how the image will look upon posting. In addition, Twitter announced the software industry's first algorithmic bias bounty competition [15], held at the DEFCON AI Village in 2021. Twitter provided their image cropping code as open source, and asked participants to submit bias assessments to help identify a broader range of issues. In September 2021, researchers from Twitter shared the results of the competition and acknowledged that it's impossible to foresee all potential issues within a professional audit, thus highlighting the importance of direct feedback from the communities using their product [48].

### 3.2 ImageNet Roulette Case

ImageNet Roulette [19] was an art exhibit publicly launched in mid-September 2019, following a physical exhibit opened in March, to expose underlying biases found within computer vision algorithms that used the ImageNet data set [20]. To use ImageNet Roulette, a user could upload an image of themselves to the project website, of their friends, or even public figures. The tool would return a modified image that flagged any detected faces and see labels describing the person in the image. The project website also had a feature enabling users to share their results on social media, including Twitter. Some users found the returned labels to be racist and sexist [62]. Several of these classification outcomes were shared on Twitter, along with the user's reactions to what they were observing.

About two weeks after its public launch, the creators of the exhibit, Trevor Paglen and Kate Crawford, felt their experiment had proved its point and retired the project website on September 27, 2019. As a result of the project, and the attention it generated, about 600,000 images were removed from ImageNet [62].

### 3.3 Portrait AI Case

Portrait AI is a website and smartphone app that, in its initial form, used AI techniques to modify an uploaded image to resemble an 18th century European painting.[3] The Android and iOS smartphone apps were released around February 2020, and the web site and apps are still active as of this writing, having since added many other art styles and visual effects. Portrait AI appears to have been created by a single developer, and the Android app has been installed over one million times.

---

[1] https://twitter.com/colinmadland/status/1307111816250748933
[2] https://twitter.com/Twitter/status/1390026628957417473
[3] https://portraitai.app/

Around May 2020, some users started to raise concerns about Portrait AI on Twitter, observing that it transformed people of color into Caucasians.[4] As of this writing, these issues have not been resolved and users still occasionally post their observations on Twitter. But, a note has been added to the app description for both iOS and Android, stating: "Note for People of Color: We are so sorry that our AI has been trained mostly on portraits of people of European ethnicity. We're planning to fix this soon."[5]

Portrait AI, unlike the previous two cases outlined above, is an existing system where many users have stumbled across possible bias, but the audits have not yet resulted in many news articles or a change in the algorithm. The majority of Portrait AI users continue to use and post their results without much mention of the biased behaviors.

### 3.4 Apple Card Case

Apple and Goldman Sachs released their new credit card, the Apple Card, on August 20, 2019 [1]. A few months later, on November 7, 2019, David Heinemeier Hansson shared his and his wife's negative experience with the Apple Card, reporting that he received a 20x higher credit limit than his wife, despite "filing joint tax returns, living in a community-property state, and having been married for a long time." He claimed that the program was sexist and was frustrated when told that no one was authorized to share the assessment process, and that it was "just the algorithm." [6] Hansson's tweet thread led to an uproar of other consumers testifying similar experiences, including Apple co-founder Steve Wozniak.[7] Later on, Apple Card manually increased their credit limit[8].

Goldman Sachs, the issuing bank for the Apple Card, responded that their process was not biased, as they did not collect or consider gender or marital status during the application process [47]. Later, the New York Division of Financial Services responded to the public upheaval on November 9, 2019 and stated that it would start their own investigation.[9] About a year and a half later, NYDFS found that there was no illegal discrimination against women and that the problematic behavior was due to poor management of the product's release [55]. Despite these findings, Twitter users continue to report disparities between their Apple Card applications and credit limits and those of others, expressing their belief Goldman Sachs' algorithm displayed gender bias. Journalist Liz O'Sullivan, of Tech Crunch, alleged, further:

"there is no doubt in my mind that the Goldman/Apple algorithm discriminates, along with every other credit scoring and underwriting algorithm on the market today. Nor do I doubt that these algorithms would fall apart if researchers were ever granted access to the models and data we would need to validate this claim. I know this because the NYDFS partially released its methodology for vetting the Goldman algorithm, and as you might expect, their audit fell far short of the standards held by modern algorithm auditors today."[10]

Similar to Portrait AI, Apple Card was a product with an underlying algorithm that users claimed to be biased. Similar to the Twitter Image Cropping case, users successfully drew enough attention that an authority (NYDFS) conducted an investigation. However, unlike all the previous cases outlined above, the Apple Card audit was sparked by noteworthy figures who had large platforms on Twitter and generated a large amount of media coverage and public attention. Examples of such noteworthy figures include David Hansson, the initiator of the audit and the creator of Ruby on Rails; John Legend, a famous musician; Steve Wozniak; and Apple Card itself. Also, much of the evidence provided in the user audit consisted of user experience reports and testimonies, unlike the large collection of images for the other cases.

## 4 METHODOLOGY

For this paper, we collected and analyzed data from Twitter because people participating in user-driven audits have used the platform to share information, raise awareness and request platform operators and authorities to respond to bias issues. Even in cases where algorithm bias occurred on other platforms, users have primarily used Twitter — and not the platform where the bias occurred — to achieve these ends [44]. Similarly, users participating in the user-driven auditing cases analyzed for this paper were sharing information about the algorithm biases on Twitter. While the algorithm biases at the heart of three cases in this paper occurred on other platforms, we did not collect data from these platforms because they did not have information sharing or social media features that could enable users to share information about the biases with each other or to raise the awareness of people who had not used the algorithms before. As such, in the following, we detail how we collected and analyzed Twitter data about the four algorithm bias cases.

*Executive Summary of Methodology:* Preceding our detailed account of the procedures taken, here, we provide a high-level summary of our methodology. To collect Twitter data about the four user-driven audit cases presented in this paper, we used the following approach. First, we collected tweets from the platform by searching for relevant keywords and then recursively collecting additional tweets referenced by the original tweets. This resulted in an initial set of 33,597 tweets. We then built a relevance classifier which filtered out irrelevant tweets, resulting in a final data set of 17,984 tweets. We then analyzed the data, examining who the most influential tweeters were as well as how different tweets played different roles in finding, discussing, and surfacing potentially harmful algorithmic behaviors. For the latter, we manually coded a subset of our entire corpus to qualitatively analyze the data by iteratively developing and refining a coding scheme. Finally, we built an ML classifier to apply our codes to our entire corpus. For privacy reasons, we redacted usernames and URLs in tweets presented in this paper, as recommended by Fiesler and Proferes [29]. The exception is with our Background section (Section 3), since in those cases the originating tweets and the names of those users were highly publicized by popular press.

---

[4] See for example https://twitter.com/blutmut/status/1263757322633052162
[5] https://apps.apple.com/us/app/portraitai-classic-portrait/id1474684190
[6] https://twitter.com/dhh/status/1192540900393705474
[7] https://twitter.com/stevewoz/status/1193424787248279552
[8] https://twitter.com/dhh/status/1192944667202998272
[9] https://twitter.com/LindaLacewell/status/1193183785581498369
[10] https://techcrunch.com/2021/08/14/how-the-law-got-it-wrong-with-apple-card/

Below, we describe these steps in detail.

## 4.1 Data Collection of Tweets

In this section, we provide specific details about our data collection method. For each of the user-driven audit cases examined, we collected data from Twitter using the following approach. First, we manually searched Twitter using the Python tool snscrape[11], which requests data via Twitter's REST API. The snscrape tool takes as input keywords and a time frame, searches for tweets with relevant phrases and hashtags, and then returns a set of URLs of relevant tweets. We iteratively developed a list of keywords, phrases, and hashtags to search on based on our knowledge of these cases as well as by reviewing popular press articles about these cases, such as "Twitter Cropping" and "Image Cropping".

To identify the time frame for each case, we examined popular press and other sources to determine a rough start-date and end-date; we used these start- and end- dates to adjust the dates input into snscrape. For each user-driven audit case, we purposely chose slightly larger time frames than what press and media sources suggested were the start and end-dates to better ensure we collected as comprehensive of a data set as possible

Since *snscrape* only returns URLs, we wrote a custom Python script to gather additional information about each tweet, in particular the tweet content, conversation ID, and number of "likes" and "retweets." We also wrote a separate custom Python script that recursively searched for and identified tweet IDs for any tweet referenced by the initial set of tweets collected. At the end of this step, we had 33,597 tweets. Note that the Twitter Cropping case was the first data set collected and done so by a different researcher, with the procedure and scripts for collecting tweet data and engagement data slightly different than described above. However, we collected the same kinds of data in the other three cases and feel that there are no substantial differences in the procedure or the data collected.

*4.1.1 Public Access to Research Software.* A copy of the *snscrape* scripts and Python notebook we used to collect, parse and create a data-frame of Twitter data about the cases is released at: https://userdrivenaudits.github.io

## 4.2 Relevance Classifier to Filter Irrelevant Tweets

In this section, we detail the procedures taken to classify whether tweets were relevant to the user-driven audit cases. In a preliminary inspection of our crawled data, we discovered a fairly large number of tweets held no relevant discussion or material relating to any of our four audit cases. For example, from the ImageNet Roulette case, one irrelevant tweet said: "Absolutely obsessed with this [URL Redacted]." Another irrelevant tweet, from the Apple Card case, said: "Why the f is that dog walking with Aborigines when he is putting them backwards with bias aboriginal shopping card saying one bad Apple all bad." To address this problem, we developed a relevance classifier to filter out likely noise.

To build this relevance classifier, we first took a sample of 1,200 tweets from our original set of 33,597. The sample included tweets from each of the four user-driven audit cases. Of these sample tweets, 500 tweets were coded as "relevant" or "irrelevant" by four coders in tandem, who simultaneously discussed and iterated on the definition of relevance. Once the 500 sample was coded, inter-rater reliability was calculated (0.848) to ensure that all coders were in relative agreement before the remainder of the sample was coded. Once it was confirmed that agreement was above 0.70 [50] the remaining 700 tweets were coded individually among the four coders.

When all 1,200 tweets were coded, we used this labeled data to train a classifier to help us filter our entire data set for noise. To build this relevance classifier, we split the hand-coded data set into training and test sets, where 70% of the data set (840 instances) was used for training and 30% of the data set (360 instances) was used for test purposes. We ran experiments and reviewed the performance of the models we tested to select the model we used to build our relevance classifier, which was a support vector machine (SVM) model. We added and used the model with the additional features of unigrams, bigrams, trigrams, and line length. Finally, we trained, tested and ran the final model using cross-validation on 10 folds. The relevance classifier's final performance had an accuracy measure of 88% and had a kappa value of 0.77. Using this final relevance classifier, we filtered out all irrelevant tweets, resulting in 17,984 relevant tweets. In 4.5, we provide a link to the open-source software used to build the Relevance classifier and report the *feature engineering* steps and parameters used in the model so that others may replicate our work.

## 4.3 Data Analysis of Top Contributors

We also wanted to investigate who were the top contributors in each of the four cases. o understand the ways through which users contributed to the audits, we classified and ranked them according to two categories: (a) *content producers*, users who produced the most number of tweets in a case; and (b) *content broadcasters*, users whose tweets generated the most engagement in terms of likes and retweets in a case. For content producers, our rationale was to use number of tweets as a proxy for the amount of labor contributed to the user-driven audit. Similarly, for content broadcasters, our rationale was to understand who helped spread awareness of the case to a larger audience of users. For our analysis, we grouped the collected tweets data by participant ID to gather all the tweets tweeted by a user, sorted users in descending order according to the total number of tweets (for content producers) or the sum of likes/retweets (for content broadcasters), and selected the top ten for each. Once top contributors were classified as content producers and content broadcasters, we then inspected the Twitter profiles of these top tweeters to further analyze their credibility and influence. We labeled the account as an *expert* if the user displayed a relevant research, STEM, or technical background [65]. We also labeled the account as an *influencer/celebrity* if the account had 1,000+ followers. In this taxonomy, a user could be both an expert and an influencer/celebrity.

We further labeled each user account as personal, news, company, bot, or other. We made these determinations by taking into consideration the account biographies and getting a general sense of post patterns. For example, in the Apple Card case, the top content producer, which we labeled as a bot, had the following in its biography: "Twitter-Bot that combines headlines of US Conservative News Outlets." This was the only bot we labeled in our data set (one author also checked if top contributors were bots using: https://botometer.osome.iu.edu).

---
[11]https://github.com/JustAnotherArchivist/snscrape

## 4.4 Data Analysis of Tweets

To analyze the dynamics of user-driven audits, we followed DeVito et al. [22] and performed a two-stage coding process on our data set. First, we randomly selected a subset of 1800 tweets to develop our coding scheme via human coders. Second, we trained a machine learning classifier on these 1800 tweets and used it to categorize our entire data set.

*4.4.1 Coding Procedure.* In the first human coding stage, we developed a mutually exclusive coding scheme that centered around the different kinds of labor users would take in an algorithm audit and how they can be split into different roles, or the division of labor (see Table 3). By division of labor, we refer to the task a user's content was achieving for an audit, such as "hypothesizing", "data collection", "escalation." The entire scheme was developed through iterative discussions among the two lead authors and two undergraduate research assistants, and was also reviewed by the research team in several large group meetings.

The coding scheme was developed through both deductive and inductive thinking. Our coding scheme was also inspired by [65], which described the lifetime of an algorithm audit as initiation, raising of awareness, hypothesizing & testing and remediation. We first applied thematic analysis [8] to the first 900 tweets, coding tweets in small blocks and holding weekly meetings to review tweets that coders had trouble classifying, refining the code book with each discussion. Once 400 tweets were coded by all the coders, an in-depth discussion was held to review tweets with low agreement, resulting in our final coding scheme. When the 900 tweets were completed, two of the four coders went on to code an additional 900 tweets, resulting in a 1800 tweet data set which would be used to train the ML classifier.

*4.4.2 Classification Procedure.* To understand users' participation across all four cases on a larger scale, next, we built a ML classifier to categorize our entire data set using the same coding scheme. The sample of 1800 tweets (see above) was split into training and test sets, with 70% of the data set (N=1,260) used for training and 30% of the data set (N=540) for testing. We ran experiments using a Naive Bayes model, a support vector machine (SVM), and a sequential minimal optimization (SMO) algorithm using 10-fold cross validation, and ultimately chose the SVM classifier for its superior performance. This SVM classifier used some additional features: unigrams, bigrams, trigrams, line length, stretchy patterns (n-grams with gaps), and included punctuation to preserve hashtags. Our final classifier had an accuracy of 88% and a kappa value of 0.849, and was used to classify the entire relevant data set.

## 4.5 Data Sets, Parameters and Software for Building the Classifiers

*Training, Test and Final Data Sets.* We used different training and test data sets to build the Relevance and Division of Labor Classifiers and produce our final data set for the analysis presented in this paper. The training and test data sets for each classifier, as well as our final data set are posted online: https://userdrivenaudits.github.io. We followed guidelines for ethical use of Twitter data based on Casey Fiesler's work, including redacting user names and URLs.

*Software:* We used the open-source software LightSide [12] to train and test the performance of the models used to build the Relevance and Division of Labor Classifiers. On our project website for this project [13], we report the parameters used to train the models and build the two classifiers. Anyone may download and use Lightside and our research data to replicate the procedures we took for this study.

## 5 FINDINGS

In this section, we present the results of our analysis. First, we present descriptive statistics about the four cases, focusing on patterns of user participation and engagement. Second, we present our analysis of the top contributors in the four cases, focusing on top content contributors (in terms of number of contributions) and top content broadcasters (in terms of spreading the audit amongst their followers). Finally, we present our analysis of the five primary roles users' tweets played in these audits, describing subdivisions of labor within each role as well as variations in terms of how these roles differ across the four user-driven auditing cases.

### 5.1 Patterns of Participation and Engagement in the Auditing Cases (RQ1)

Table 1 presents descriptive statistics of user participation and engagement in each of the four user-driven audit cases. In summary, we found: (1) three out of our four cases (Portrait AI being the exception) shared a similar pattern of user participation, where there was a large initial burst of tweeting activity, followed by a long tail of low activity; (2) the vast majority of participants in all four cases contributed just a single tweet.

Of the four cases, the Twitter Cropping case had the most total tweets contributed by users (11,323 tweets), followed by ImageNet Roulette (2,813), Apple Card (2,408), and Portrait AI (1,440). Second, the most tweets posted per day by users, e.g. the "peak" participation day, occurred in a day for the Twitter Cropping case, within a week for Apple Card, around 6 months for ImageNet Roulette, and over a year for Portrait AI. Note that for ImageNet Roulette, the first tweet we found about it was from about 6 months before its public launch date when it was still a physical exhibit. As such, the peak for ImageNet Roulette actually represents its public launch date.

Next, we observed three of our four cases – the Twitter Image Cropping, ImageNet Roulette, and Apple Card – had a similar pattern of user participation, where there was a large initial burst of tweets followed by a long tail of low user activity (see Figure 1). One possible explanation for this pattern is these respective audits engendered a great deal of attention from celebrities, public figures, and other influencers, which led to interventions by institutions or technology platform operators who then took actions to address the concerns users raised. For instance, in the Twitter Image Cropping case, the peak of user participation was just 3 days after a user posted the initial tweet pointing out the problem with the algorithm. The same day this first tweet was posted, Twitter apologized and outlined steps it would start taking to address the algorithm's racial bias problem [41]. Similarly, for Apple Card, the New York Department of Financial Services reported on November 9, 2019 it would start investigating the Goldman Sach's creditworthiness algorithm, just a few days after the original tweet was posted by a user alleging that the algorithm was biased [69].

---
[12] http://ankara.lti.cs.cmu.edu/side/download.html
[13] https://userdrivenaudits.github.io

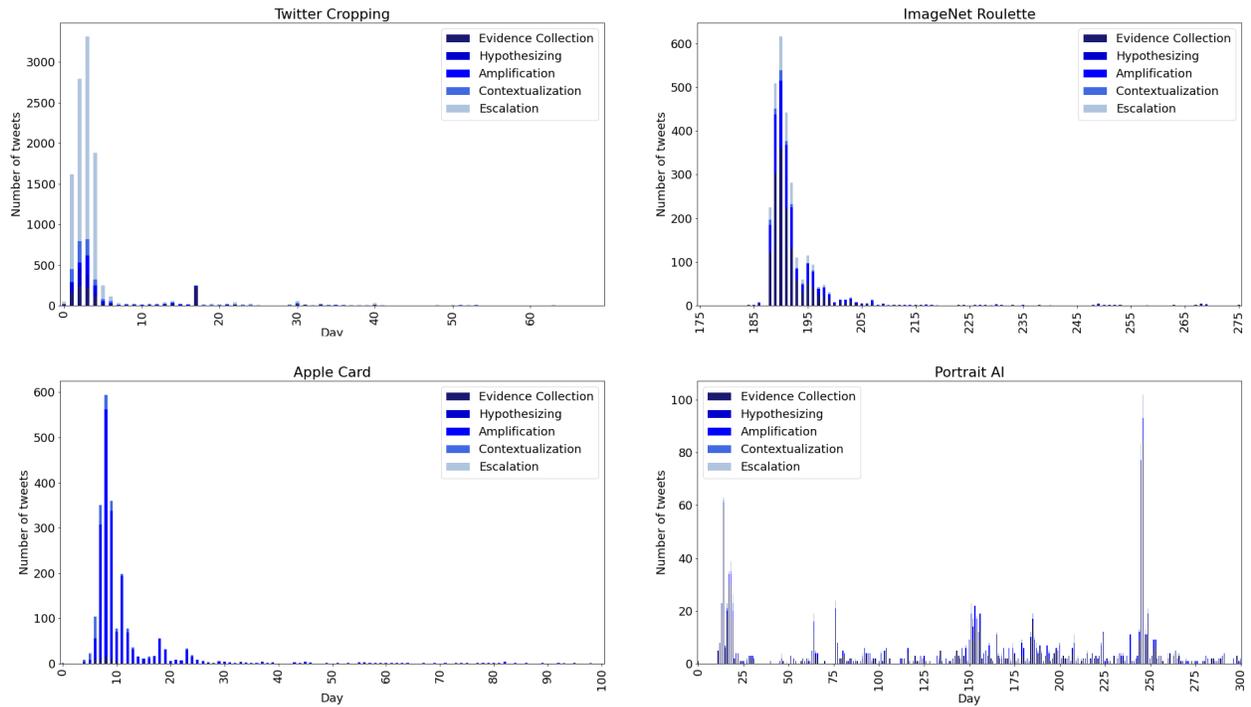

Figure 1: Tweet activity by Division of Labor/Role over time. In all of the cases, the x-axis has been set so that 0 is the day of the first tweet that originated the user-driven audit. The stacked colors of the bars represent the percentage of tweets playing one of the 5 'Division of Labor' roles (these are listed in the legend). Note that the x-axes and the y-axes differ for each case. Also, for ImageNet Roulette, the first tweet about it was about 6 months before its public launch date. As such, due to page size and fit, the X-axis starts at Day 175 in the case and the figure represents the case's peak activity, which occured after the art project's public launch. In 5.3 the major user auditing roles displayed in these figures are described.

Table 1: Comparing Descriptive Statistics of our Four Cases. In this paper, a conversation is defined as either: (1) a set of tweets that share a "conversation_id" or (2) as a tweet, and its corresponding replies and the replies to those replies. User contribution is defined as the number of tweets associated with a user-driven audit posted to Twitter by a participant.

|  | **Twitter Cropping** | **ImageNet Roulette** | **Portrait AI** | **Apple Card** |
| --- | --- | --- | --- | --- |
| Date Range | 09/18/20 - 11/26/20 | 03/12/19 - 05/13/21 | 05/08/20 - 01/12/22 | 11/03/19 - 01/16/22 |
| Duration | 2mth 8dys | 2yrs 2mth 1dys | 1yr 8mth 4dys | 2yrs 2mth 5dys |
| Total # of Tweets | 11,323 | 2,813 | 1,440 | 2,408 |
| Most Tweets in One Day | 3,378 | 617 | 102 | 595 |
| Highest Peak Date | 09/21/20 | 09/18/19 | 01/09/21 | 11/11/19 |
| # of Unique Participants | 10,160 | 2,286 | 1,226 | 1,687 |
| # of Conversations | 883 | 2,233 | 1,335 | 2,071 |
| Avg. Tweets per Participant (stdev) | 1.11 (2.41) | 1.23 (0.78) | 1.17 (0.73) | 1.43 (2.49) |
| Median | 1 | 1 | 1 | 1 |
| Avg. Tweets per Conversation (stdev) | 12.82 (129.40) | 1.26 (1.94) | 1.08 (0.37) | 1.16 (4.15) |
| Median | 1 | 1 | 1 | 1 |
| Avg. Participants per Conv. (stdev) | 11.77 (124.46) | 1.14 (1.72) | 1.04 (0.21) | 1.08 (2.11) |
| Median | 1 | 1 | 1 | 1 |
| % Participants with > 1 tweets | 5.87% | 13.69% | 10.28 % | 12.66% |
| Total # of Likes | 695,094 | 46,698 | 46,429 | 200,681 |
| Total # of Retweets | 10,926 | 3,807 | 1,503 | 8,537 |

In contrast, the Portrait AI case showed sporadic conversation and user tweeting activity throughout the observed time frame for the case. Compared to the other user-driven audits, only a few users raised concerns since the software's inception about whether its algorithm was biased. Furthermore, the platform operator has not yet directly addressed the hypothesized bias issue, nor has any other authority intervened. We note that for the Portrait AI case, there have been very few news articles about potential bias (unlike the other three cases), nor have there been any celebrities discussing it (unlike the Apple Card case).

Last, we found the vast majority of participants in all four cases contributed just a single tweet. While all cases had many participants, very few of them had significant participation in terms of number of tweets. Furthermore, while any one of these single tweets likely had little impact, in aggregate we believe they increased the profile of these cases and drew the attention of journalists, which ultimately led to changes in the Twitter Image Cropping, ImageNet Roulette, and Apple Card cases.

## 5.2 Top Contributors: Who They Are and What They Do (RQ2)

In this section, we present the results of our analysis of the top contributors for each case, focusing on "content producers" and "content broadcasters". As described in Section 4, content producers produced the most tweets in a case (using this as a proxy for contribution towards the user-driven audit), and content broadcasters had the highest engagement metrics of likes and retweets (using these as a proxy for spreading awareness of the case). We found that, among the tweeters who actively contributed to these audits, there was a clear division between "top content producers" (people who produced and posted the largest number of tweets) and "top broadcasters" (people who received a large number of likes and retweets). In other words, although top content producers might have contributed more tweets, their content garnered less engagement than top broadcasters, suggesting user-led audits may require both types of users to be successful.

For example, with the ImageNet Roulette, Portrait AI, and Apple Card cases, there was only one account for each case that was both a "top 10 producer" and "top 10 broadcaster". Furthermore, for the ImageNet Roulette and Apple Card cases, the overlapping accounts also happened to be the initiators of the two cases. Kate Crawford of the ImageNet Roulette case was ranked as the second-highest content producer and content broadcaster, while David Heinemeier Hansson was ranked as the second-highest content producer and the highest-ranked content broadcaster.

We also observed a difference in the number of influencers in top content producers and top content broadcasters. We defined "influencers" as users who have a current follower account greater than 1,000 at the time of our analysis (note that this might not be the same as at the time the user participated in the case). Top content broadcasters had a higher number of influencers than top content producers. Because content broadcasters are characterized by the number of likes and retweets their tweets received, it comes as no surprise that users with large followings on Twitter would become top content broadcasters. This could indicate that having the attention and contribution of influencers can help the spread of a user-driven audit case.

## 5.3 The Division of Labor in User-Generated Content (RQ3)

Our analysis of tweets led to five roles that tweets play in the auditing cases: hypothesizing, evidence collection, amplification, contextualization, and escalation. Note we treated all tweets associated with a case as a type of 'harm identification' since these audits were raising attention about algorithm bias (please see 4.2 for details on how we determined the relevance of tweets).

Table 3 shows the description and prevalence of each role in the four audit cases. All the roles existed across all the auditing cases; however, their prevalence varied from case to case. For example, the majority of tweets in the Twitter Cropping played the role of escalation by providing emotional reactions to what has been observed within the audit, while escalation tweets were almost negligible in the Apple Card case. We discuss these patterns in detail when describing each role below.

To understand the distribution of the tweet roles across the lifetime of each case, Figure 1 illustrates stacked bar charts to display what percent of tweet activity on a given day was playing each role. This shows that the participants and roles are intertwined, corroborating previous work that user-driven algorithm auditing is not necessarily a linear/staged process [65]. For example, we observed that for the Twitter Cropping, ImageNet Roulette, Portrait AI and Apple Card cases users tweets played each of the roles over time, even in cases where one role dominated. For these cases, therefore, our data does not suggest users played different roles at different times or that they played certain roles early in the case but not later.

*5.3.1 Hypothesizing.* The first category of tweets is *hypothesizing*, in which users proposed different informal theories they developed about algorithmic systems [22, 25] to explain how the algorithms operated to produce these harmful results [34]. In general, we observed two kinds of hypothesizing tweets: theoretical hypothesizing and experimental hypothesizing. Across the four cases, "hypothesizing" tweets made up one of the smallest portions within the respective data set (refer to Table 3).

*Theoretical Hypothesizing:* In theoretical hypothesizing tweets, users stated what they believed caused the algorithm's bias(es). Users' beliefs about the sources of the bias were seemingly informed by their occupation or work experience, knowledge, and awareness of how algorithms generally make predictions or decisions. For example, one participant believed data set homogeneity explained why ImageNet Roulette produced seemingly biased image classifications: *"Should this be a surprise? No. Most #AI is based* on statistical merging of data from many people. Just like a corporation. We need approaches to AI that reflect the diversity of people, not burying it in big data. [URL redacted]" (T14034).

Table 2: Comparing top ten content producers and content broadcasters. Please note that number of kinds of accounts may not add up to 10, as some accounts have been made private and deleted over the course of our analysis.

|  | Twitter Cropping | | ImageNet Roulette | | Portrait AI | | Apple Card | |
| --- | --- | --- | --- | --- | --- | --- | --- | --- |
|  | Producer | Broadcaster | Producer | Broadcaster | Producer | Broadcaster | Producer | Broadcaster |
| # Experts | 7 | 5 | 6 | 3 | 1 | 2 | 4 | 7 |
| # Influencers/Celebrities | 4 | 8 | 7 | 9 | 4 | 7 | 8 | 10 |
| # Personal Accounts | 9 | 10 | 8 | 9 | 10 | 7 | 7 | 7 |
| # News Accounts | 0 | 0 | 0 | 0 | 0 | 0 | 2 | 1 |
| # Company Accounts | 0 | 0 | 0 | 0 | 0 | 1 | 0 | 1 |
| # Bots | 0 | 0 | 0 | 0 | 0 | 0 | 1 | 0 |
| # Other | 1 | 0 | 1 | 0 | 0 | 0 | 0 | 0 |
| Avg. #Followers (Top10) | 7,068.8 | 42,460 | 31,026.9 | 336,700.2 | 4,078.7 | 9,115.9 | 892,758.8 | 2,133,274 |
| Median | 839 | 7,374 | 11,149 | 421,517 | 587.5 | 4,718 | 2,720.5 | 123,054 |
| Avg. #Followers (Not Top) | 95.5 | 229.5 | 216.9 | 236.8 | 264.1 | 109.3 | 211.4 | 156 |
| Median | 58 | 127.5 | 66.5 | 315.5 | 179 | 129.5 | 87.5 | 129.5 |

Table 3: Division of Labor codes. These codes describe the work done in an audit. A tweet will have at most one of these tags. Specific tweets in our data set are labeled as T####, e.g. T2538. Percentage represents the portion of tweets with this tag across our four cases after applying our ML model: Twitter Image Cropping (TC), ImageNet Roulette (INR), Portrait AI (PAI), Apple Card (AC).

| Code | Definition | Example | Percentage by Case |
| --- | --- | --- | --- |
| Hypothesizing | Tweets which hypothesize why the observed behavior of the algorithm is happening. | "@XXXXX Not that it isn't problematic, but I think what's happening here is the algorithm prefers the higher contrast between facial features and brightness of skin tone. When converted into binary code, the algorithm is more confident that the light-skin face is an actual face." (T2538) | TC: 5.69% INR: 0.14% PAI: 0.49% AC: 0.29% |
| Evidence Collection | Tweets which include evidence that support or oppose a hypothesis proposed by other tweets Typically includes a link to an image, or in the Apple Card case which a user testimony/experience. | "i got typed as 'clown, buffoon, goof, goofball, merry andrew' on imagenet roulette... it's like they've known me all my life [URL Redacted]" (T13072) | TC: 9.68% INR: 52.44% PAI: 72.5% AC: 3.65% |
| Amplification | Tweets which help share and broadcast information necessary for others to participant in the audit and to raise awareness of findings and relevant discussions. Typically includes tagging other Twitter users and/or sharing news articles about the case. | "This is a good article about the whole stink @XXXXX made about Apple Card's apparent bias against women. The original thread comes across as whiny and super annoying, but there are some worthwhile takeaways from the situation. [URL Redacted]" (T16514) | TC: 4.66% INR: 30.61% PAI: 17.57% AC: 85.05% |
| Contextualization | Tweets that contextualize the ongoing auditing and placing it within the broader social, technical or cultural context. | "@XXX When techbros try to tell you why they don't need to take an ethics class" (T10823) | TC: 8.09% INR: 3.34% PAI: 1.11% AC: 10.63% |
| Escalation | Typically emotional tweets that primarily react to what's been observed within the audit, both implicitly increasing the visibility of the case and building a counter-public sphere among auditors. | "Bruuuuuhhhh why PortraitAI do me like that?? [URL Redacted]" (T14284) | TC: 69.21% INR: 13.30% PAI: 8.13% AC: 0.33% |

We also observed users offering hypotheses to counter others' beliefs. For example, another user countered the data set bias above, instead proposing users lacked understanding that ImageNet Roulette was designed as an art exhibit to showcase how algorithms were biased, and this best explained why users observed bias: *"Thought-provoking essay on the political issues of image recognition research: its not just about better datasets. (ImageNetRoulette was over-simplified/-sold in the media; I think it's a better project when you understand that how it is exaggerating a real problem for effect.) [URL redacted]" (T13876)*. Some also expresses their belief that algorithmic bias stems from the biased people behind the algorithm, including developers and researchers, not that algorithm itself: *"The only ML algorithms with race and gender biases are those concocted by "ethical AI" researchers." (T17807)*.

*Experimental Hypothesizing:* Experimental hypothesis tweets were characterized by users offering testing strategies or actions that others could employ to identify and evidence if an algorithm was biased. In other words, these tweets suggested further courses of action for users to adopt to support a hypothesis. For example, in response to one user's tests, another participant suggested additional tests that could also be used evaluate Twitter's auto-cropping algorithm. *"@bascule Now try Obama in the first picture on top and Obama in the last picture at the bottom. Maybe it's just how twitter organizes pictures first top last bottom." (T2883).*

An aspect of these hypotheses we found was that users proposed different theories, and they were influencing each other. The theories proposed would include and/or remix aspects of others' beliefs about how the algorithm was working. For example, in the Twitter Image Cropping case, while some users hypothesized skin color or race was the reason for cutting out an image, some others proposed checking for other facial features by switching these features in images and seeing the results: *"@XXXXX A better experiment would be to artificially darken Mitch's face and whiten Obama's face and have a black Mitch vs white Mitch or black Obama vs white Obama test. As others pointed out it could be other facial features that affect the algorithm selection, like smile or glasses." (T2212).*

5.3.2 **Evidence Collection**. The second category of content was *evidence collection* tweets, where users shared evidence they had collected to document a software algorithm's bias(es). Their evidence typically supported hypotheses proposed by other users and some occasionally invalidated one of these hypotheses. In other words, with the evidence users collected, the auditing cases produced digital evidence for the claims made by other participants. It is also possible that evidence collection tweets laid the groundwork needed to garner the attention of authority figures, such as in the Apple Card case when the New York Department of Financial Services responded on Twitter [14].

Across all four cases, evidence collection tweets represented a substantial portion of the respective data sets, e.g. making up 52.44% of the ImageNet Roulette case and 72.5% of the Portrait AI case. The ImageNet Roulette case features a top-down user-driven audit, where the biased behavior has already been publicly established which could potentially influence participants to want to test and see the biased behavior themselves, possibly resulting in a larger volume of "evidence collection" tweets. The Portrait AI case is interesting here because many people often shared the results of their portrait without explicitly trying to audit the system.

For the Twitter Image Cropping, ImageNet Roulette, and Portrait AI cases, we found user-gathered data were typically altered images users produced, gathered and shared on Twitter by uploading an original image to a software app whose algorithm then transformed the image data in some way. For example, in Tweet 15, a user described A/B testing images to discern if the alleged algorithm bias in the Twitter Image Cropping case was observable: *"OK, so I am conducting a systematic experiment to see if the cropping bias is real. I am programmatically tweeting (using tweepy) a (3 x 1) image grid consisting of a self identified Black-Male + blank image + self identified White-Male ( h/t @XXXXX @XXXXX)" (T15).*

In the Apple Card case though, the user-generated data or evidence was constituted largely of user-generated text content, e.g. tweet text, describing their testimonies of applying for an Apple Card. These tweets noted their observations of disparate decisions that Goldman Sach's algorithm had made about their or others' credit approvals or limits, mainly alleging that the algorithm was gender biased: *"Fascinating. @Apple still has bias. My wife and I each applied separately for the Apple Card. Identical income and credit scores (hers is higher). I got a 25% higher limit. @XXXXX" (T15654).* Another user in T17118, also accused Apple and Goldman Sach's algorithm of racial bias, saying the algorithm determined they had a lower creditworthiness rating than what other credit bureaus had determined: *"It's racially biased as well. Stated my FICO was 100 points LESS than the other major reporting bureau #AppleCard Apple Card faces scrutiny following allegations of gender bias [URL Redacted]" (T17118).*

*Types of Media Data Collected and User Comparisons of the Data:* Across three cases, all except Apple Card, we found users typically either tested and shared images of others or themselves. For the Twitter Image Cropping case, we found users commonly posted images of celebrities, public figures or politicians when testing the platform's image cropping algorithm. In comparison, users engaged in evidence collection activities for ImageNet Roulette and Portrait AI often tested and shared images of themselves through the algorithms of these software. We observed this meant users collecting data for the Twitter Image Cropping case were producing and sharing images that were potentially more comparable. For example, many users tested and shared their testing results for cropped images of President Barack Obama and U.S. Senator Mitch McConnell. In comparison, ImageNet Roulette and Portrait AI users usually tested and shared classified images of themselves, typically a portrait of their face.

Once data collectors posted images to Twitter, we found users often compared and discussed their observations and beliefs about the pictures, sometimes proposing new hypotheses to explain why a software's algorithm classified or cropped the images in the given manner.

5.3.3 **Amplification**. The third category of tweets was *amplification*, in which users broadcast and amplify relevant information related to the audit. These tweets either include information and resources themselves, such as findings from other users, or provide direction towards relevant resources to support newcomers. When users post amplification tweets, they expose their followers to the ongoing audit, which increases the visibility of the case and potentially attracts others to take part in the audit themselves. Amplification tweets took up a large portion of ImageNet Roulette (30.61%), Portrait AI (17.57%), and especially Apple Card (85.05%). The Apple Card case had received a lot of media coverage from beginning to end, meaning there were a lot of articles that could potentially be shared. This could possibly explain why the vast majority of the Apple Card tweets were coded as amplification. ImageNet Roulette also garnered some media attention, potentially due to the influence of Kate Crawford and Trevor Paglen, the creators of the project. The ImageNet Roulette case had a large portion of evidence collection tweets, which could potentially prompt other Twitter users to try it themselves.

In each case, we believe there was a distinct demand for some users to explain and provide information about an audit to others. As seen with the amplification tweets across cases, this role is distinct from tweets sharing the data and evidence users collected, as users collecting data did not always connect the evidence to the broader narrative of the audit, as well as related societal issues of bias. From our observations, we believe the amplification tweets in these cases suggested users want this type of information too. In turn, we speculate amplification tweets could trigger users playing other roles to produce more content, for example, by collecting additional data. In other words, users tweeting amplification tweets could encourage other users to adopt different roles in these audits.

---

[14] https://twitter.com/LindaLacewell/status/1193183785581498369

Ultimately, we observed amplification tweets made up the smallest portion of the Twitter Cropping case, despite having the largest volume of tweets in total. Though there may be many reasons for this observation, one possible explanation could be that the escalation tweets (described below) within the Twitter Cropping case, making up 69.21% of the data set, may generate the same effect that amplification tweets hope to accomplish. Below, we further identified three types of amplification tweets: inviting newcomers, linking news articles, and providing information and resources.

(1) Some users tag specific people, manually adding them to the discussion and pointing a potential new participant towards information about the audit. For example, in T3116, the user tags two other Twitter users to look at a thread that curates examples of Twitter's auto-cropping algorithm: *"@XXXXX @XXXXX @XXXXX must see this!" (T3116)*.

(2) Some users link news articles covering the progress of the audit. By doing so, they amplify the reach and exposure of relevant news articles. For example, in response to the events of the Twitter Image Cropping case, Twitter released a public apology for its algorithm, and the newspaper The Guardian [41] covered it in an article that was tweeted out multiple times within the audit: *"Twitter apologises for 'racist' image-cropping algorithm [URL redacted]" (T382)*.

Similarly, the ImageNet Roulette case had it's own fair share of public press, such as in Frieze Magazine [32]: *"I wrote for @frieze_magazine about @katecrawford and @trevorpaglen's #imagenetroulette [URL redacted]" (T13194)*.

And in April of 2021, when Apple Card released a new feature in response [53] to the accusations of gender bias spread on Twitter a few months prior, users included this update into the audit effort: *"Apple Card's new feature aims to address the gender bias that's all too common in the credit industry.[URL redacted]" (T15657)*.

(3) Some users include information and resources in the context of answering the question of others. By doing so, theses users point other users who have expressed interest in the events surrounding the audit towards relevant information and resources to help them get involved, if they so choose. For example, a Twitter user asks *"@XXXXX What is this site I want to do it" (T12044)*, and in response, the participant introduced ImageNet Roulette: *"@XXXXX Imagenet roulette" (T12048)*.

*5.3.4* **Contextualization**. The fourth category of tweets was *contexualization*, in which users place the ongoing audit into a larger social, technical, or cultural context, broadening the scope of what they're observing. They do so by providing the general social, technical, and cultural information that may be helpful for others to better understand the underlying algorithms, the problematic machine behaviors, and why and how they might harm other social groups. By doing so, those tweets help to inform and direct other users who might not have the social background, technical expertise, or cultural knowledge necessary to effectively understand and particiate in the audit.

Users who post contextualization tweets might criticize the user-driven audit itself, placing the efforts of other participants within a larger social and cultural context. For example, in the below tweet, a participant points out that other participants were creating and testing hypotheses despite a lack of technical knowledge, and attributes this behavior to human nature: *"@XXXXX Everyone in this thread is hopelessly probing an algorithm they don't understand, and coming up with imperfect theories based on experiment. Dope, thats what humans do. interesting scroll." (T3212)*.

Similarly, the following user throws doubt against the claims of bias within the Apple Card algorithm, stating that the audit participants were jumping to conclusions based on little evidence: *"@XXXXX This is a great example of someone immediately playing the "censorship! Bias!" card based on very limited evidence...when (if you think about it) Apple would have no reason to do this. Anyone can find DC any # of other ways...all this would do is make people mad at Apple" (T15723)*.

Contextualization tweets may also include discussion of the larger landscape of algorithmic bias which provides the technical context surrounding the audit. As seen below, a participant comments on the relevancy of bias in artificial intelligence and where the ImageNet Roulette stands within it: *"@XXXXX Bias in AI is a very interesting and relevant topic that will continue to gain importance as we move forward. Projects like #ImageNetRoulette will hopefully raise awareness around this issue and incite change. #BCSTT" (T11357)*.

In a similar fashion, the following participant within the Portrait AI user-driven audit comments on the possible technical and cultural influences that may have affected the development of the Portrait AI algorithm: *"@XXXXX @XXXXX @XXXXX Sadly there not that much classic portraits of PoC. Thats sucks but i don't think that this is deva fault, they're probably took giant collection of portraits of classic artists and just trained AI with it" (T15566)*.

Contextualization tweets made up the second largest portion of the Apple Card case (10.63%), but doesn't make up large portions of any of the other cases. Regardless, it isn't the smallest portion of tweets for any of the four audit cases. These observations could indicate that making content with this content is not popular or that only a small percentage of participants are willing / able to create tweets with this role. Since "contextualization" tweets require some level of expertise, it may be the case that these kinds of tweets are being created by a small portion of the participants.

*5.3.5* **Escalation**. The final category of tweets is *escalation*. We found that a large portion of tweets are either reactions to what they've observed in their own usage of the algorithm or what is being presented by others. Those tweets are emotional and reactive, expressing what the users are feeling about the observed algorithmic behaviors or the ongoing audit. Although escalation tweets might not contribute direct evidence to the ongoing audit or help form hypotheses, they might have nevertheless increased the visibility of the audit on algorithmic mediated platform like Twitter, in turn, increasing traffic and drawing more users towards the user-driven audit. By doing so, these tweets also help build a counter-public sphere [24] among the auditors, as individual participants are now, possibly, able to emotionally relate to one another on a common topic and/or opinion, which can create a feeling of community around the audit. These tweets could also grow the number of users who engage with users tweeting about the audits, as reactive user-generated content is known to do this [14].

The composition of escalation tweets was similar across all four cases. Escalation tweets are usually short, use acronyms, expletives, slang, include emojis, and often incorporate humor. Potentially, those elements helped escalate the emotional and subjective influence within the audit. For example, in the below tweet, the user makes another aware of what the Twitter Image Cropping algorithm did to their post, expressing their frustration with capital letters and expletives: *"@XXXXX @XXXXX TWITTER CROPPING FUCKED U OVER" (T7450)*.

In a similar fashion, a participant in the ImageNet Roulette case makes known what they thought of how the ImageNet Roulette classified their images: *"that imagenet roulette shit got me fucked up yo" (T12894)*. As another example, a participant in the Apple Card expressed their frustration at what they observed from the Apple Card algorithm: *"@XXXXX @dhh @AppleCard it's really fucked this shit goes on behind the scenes. like why tho." (T17973)*. In contrast, other users communicated their opinions using humor: *"Bruh why use PortraitAI when you can just paint yourself" (T14276)*.

Escalation tweets were the largest portion within the Twitter Cropping case (69.21%), but were relatively small for the other three cases, though it is not entirely clear why this difference exists. Also, in some cases, escalation tweets can serve as amplification, helping to express negative reactions to an algorithm and increasing the visibility of the case.

## 6 LIMITATIONS

Our work has several limitations. In this section, we describe these limitations to highlight where readers should apply caution in interpreting our results. We strongly encourage further work to be pursued within the space of user-driven algorithm auditing to build and improve upon the limitations outlined below.

First, we only collected and examined data about user-driven audits from Twitter. User-driven audits on other platforms may have different characteristics. Second, we collected our data by searching Twitter using specific key words and phrases and then using Twitter's API to automatically download data for URLs using *snscrape*. This keyword-centric approach may have missed some relevant tweets. However, using keyword searches to collect data from Twitter is a common approach to study user activities on Twitter [46], including for studies about information-sharing by users on Twitter about user-driven audits on other platforms [44].

Another limitation of our study was that we only examined 4 user-driven auditing cases on Twitter, so our results might not generalize to other types of user-driven audits on the platform. For example, the type of bias and/or the type of output an algorithm generates, the suitability of Twitter itself for user-driven audits, as well as popular press, could affect who and how users participate.We also only examined the text of tweets. For example, for the Twitter Cropping case, we did not analyze images that people used to A/B test Twitter's algorithm for racial bias. Further, some of our Twitter data was possibly produced and amplified via bots. While we did use a tool called Botometer (https://botometer.osome.iu.edu/) to evaluate if top content producers and broadcaster accounts were bots, we did not use the tool on our entire data-set. To the extent that a Twitter account was a bot and sharing relevant information associated with a user-driven audit and included in our data set, it did not necessarily make sense to exclude the account from our analysis, as such a hypothetical account could fit the roles we describe. Finally, another important technical limitation of our work is we present descriptive statistics about participation levels in user-driven audits on Twitter. These descriptive results cannot definitively identify any causes or reasons for these levels.

In addition to these technical considerations, another limitation of our study stemmed from our position as researchers. Our interpretation of the role a tweet played could differ from both the intended message of the author and Twitter audience. Also, our work addressed issues of algorithm bias impacting communities to which our research team do not belong, for example Black Twitter users. Our work could represent or introduce its own biases in ways harmful to marginalized demographics.

## 7 DISCUSSION AND DESIGN IMPLICATIONS

In this paper, we offered a comparative analysis of four highly visible cases of user-driven algorithm auditing on Twitter. Based on our findings, we now discuss what is "success" in a user-driven algorithm audit, how to think about users' participation in user-driven audits as well as potential design implications to better support such audits in the future.

### 7.1 What is "Success" for User-Driven Algorithm Audits?

One question that arises from our work is what success means for user-driven algorithm audits. Previous literature on algorithm audits has proposed certain criteria to determine the success of an audit. For example, Costanza-Chock et al [17] state: "An AI auditor evaluates according to a specific set of criteria and provides findings and recommendations to the auditee, to the public, and/or to another actor, such as to a regulatory agency or as evidence in a legal proceeding." However, these criteria do not capture many aspects of the user-driven audits we examined, despite the fact that at least three cases we examined (Twitter photo cropping, ImageNet Roulette, and Apple Card) could be considered successful in that they led to a significant intervention. For example, the user-driven audits we examined did not have a clear end-goal. The tweets initiating each respective case proposed a hypothesis as to why the behavior happened. However, there is no evidence that the respective authors of these initiating tweets expected others to collect data, test their hypotheses, and come up with other competing hypotheses.

Unlike conventional first-, second- and third-party audits [17], the success of a user-driven audit, therefore, might no longer be solely determined by whether there is a clear set of solutions for fixing the problem. Indeed, among all the four user-driven audits examined in this paper, none of them had a specific set of criteria or recommendations for any entity. There was no appearance of these audits trying to be comprehensive or complete, with much of the data gathering and hypothesizing focused on a narrow aspect of the respective algorithm rather than being comprehensive. Sometimes, there was not even a clear desired outcome. Below, we describe a series of different perspectives and criteria that can re-define success in the user-driven algorithm audits, and inform design implications of developing platforms and processes to achieve "success."

*7.1.1* **Centering "Ethics of Care"**. One useful framework for thinking about user-driven audits is feminist ethicist Carol Gilligan's differentiation of "Ethics of Justice" from "Ethics of Care" [35]. Ethics of Justice is usually preferred by dominant social groups, and focuses more on generalizable standards and solutions. In contrast, Ethics of Care centers more on interpersonal relationships and care as a virtue. Similar to what Wu et al. [73] found among participants in a study about collective action for privacy, in our four cases, users engaging in algorithm collective audits are also more leaning towards "Ethics of Care" instead of "Ethics of Justice". Indeed, instead of scientific tests, hypothesis, and measurements, all the cases we examined here were composed by a large number of tweets focusing on emotional expression (i.e., "escalation" and "amplification"). In other words, viewing via the lens of "Ethics of Care," a successful user-driven audit might lack clear, universal rules or solutions, but can instead center more on relationships, responsibilities, and emotional expressions. Therefore, instead of channeling all the users' efforts towards scientific hypothesizing and testing, the design of future user-driven auditing platforms or processes should also support emotional expression, communication, and community building efforts.

*7.1.2* **Enabling resistance in an incidental, fluid form**. Another way of thinking about the success in user-driven audits is to consider it as a form of "everyday resistance" [65]. Instead of thinking of resistance as highly organized actions that pose a revolutionary challenge, this line of research foregrounds a different type of resistance that is more incidental but nevertheless continuously contesting the existing power structures in everyday life [64]. In contrast to conventional audits, the success of a user-driven audit, therefore, cannot be solely determined by whether it has a highly organized structure. As we saw in our four cases, the participants involved in the user-driven audit could come and go as they please, contributing as much or as little as they desired. The audits were also highly unstructured, with many people testing things on their own initiative. Three of the audits (Twitter Image Cropping, Apple Card, and ImageNet Roulette) were rather short, with the bulk of the tweets for each case made in a small window of just a few days. However, such highly spontaneous and fluid organizational structure has also given users enough freedom and flexibility to participate based on their own comfort level, time, and availability (e.g., transportation, Internet access, childcare, etc.) in their everyday lives [59]. For designers, this indicates a need to carefully design scaffolding techniques to guide the users, while also keeping some level of spontaneity and flexibility to provide them with a wide range of choices for participation.

*7.1.3* **Forming mini-counterpublics**. Success in user-driven audits can also be viewed via the lens of "counterpublics" [31], that is, as a community building effort. As we saw across all the four cases, although the majority of the users only contributed a small amount of tweets, they have nevertheless formed small, temporal online communities around emerging problematic machine behaviors. We may consider such communities as what Nancy Fraser termed as "counterpublics" [31] – where members of often marginalized social groups collectively participate in their own form of sensemaking, opinion formation, and consensus building. A success in user-driven auditing, therefore, also means that users are coming together to build mini counter-publics, as part of a community building effort. This calls for designing and building affordances in user-driven algorithm auditing platforms and processes that enable building such counterpublics by aiding users to come together, discuss, and build community.

*7.1.4* **Raising Awareness**. Finally, unlike other conventional audits whose ultimate goal is to offer clear-cut *technical* remediation, we found "awareness raising" to be a critical dimension of success in user-driven audits [65]. Recent literature suggests that in cases where a system developer does not want to host a user auditing tool on the platform, "end-user audits would primarily seek to effect change by naming and drawing attention to problematic system behavior" [49]. Indeed, although all the four cases we examined here did not have clear set of solutions or well-structured action plan, they have nevertheless resulted in different level of changes, via the process of "awareness raising" – publicly sharing their hypotheses and evidence, actively boosting the visibility of the audits via amplification and escalation. With the Twitter Image Cropping case, Twitter quickly acknowledged the failure, explained how they tested their algorithm and shortcomings in their method, and stated some next steps to address the problem [3]. With the Apple Card case, the New York Division of Financial Services quickly stated that they would investigate [69]. With ImageNet Roulette, while there was not immediate action, egregious labels were eventually removed from the data set [62].

## 7.2 Participation in User-Driven Algorithm Audits: Activism or Slacktivisim?

As previous work discussed, algorithm audits in many cases can be understood as a form of activism, "a practice of direct action, often with the effect of drawing attention to an issue and bringing about political and social change" [52]. However, in this study, the vast majority of participants only contributed one tweet, most often in the form of an "Escalation" tweet. Some may depict such behavior as "slacktivism," that is, a "low-risk, low-cost activity [...], whose purpose is to raise awareness, produce change, or grant satisfaction to the person engaged in the activity" [61]. And such behavior is often viewed in contrast to "practical activism" – "the use of a direct, proactive and often confrontational action towards attaining a societal change" [61]. So the question that arises here is: How should we understand the role of these "slacktivists" in a user-driven algorithm audit?

We argue that, while in a more conventional algorithm audit, whose goal is to uncover biases in an algorithmic system in a systematic way, such "slacktivists" and their single tweet might look unhelpful, having a large number of slacktivist tweets in a user-driven audit can be actually impactful. One the one hand, their tweets demonstrated a high level of surprise or outrage over a perceived harmful behavior. On the other hand, their tweets also helpied raise awareness of the case (which is discussed in previous work as well [51, 61]), by sharing it with followers and algorithmically boosting the case onto Twitter's trending topics. We also believe that these two factors helped raise enough concerns to get the attention of other entities who can further help raise awareness (e.g. popular press) or intervene directly into the harmful machine behaviors (e.g. regulatory agencies or the company using the algorithm). This aligns with Cabrera et al.'s argument that "sometimes activism, in a digital age, relies on the slacktivism of the masses" [12]. But how do slacktivists get involved in the auditing process for pushing for a change?

We observed that "influencers" and "experts" were among the top contributors in the auditing cases who played a critical role in drawing attention to the problem as well as potentially attracting more people to participate. Influencers were perhaps most notable with the Apple Card case, where some celebrities amplified the case with their followers. It is likely that influencers are essential for a successful user-driven audit, to greatly raise awareness and to help recruit potential participants. In addition, in many cases, "experts" – many of them were involved as content producers and content broadcasters – also actively steered the direction of the audits by offering hypotheses, counter hypotheses, contextualization, and data. This might be partly due to a confluence of interest in the problem as well as technical ability or data analysis skills. Therefore, while most of the user-driven audits were not led by centralized organizers, these "influencers" and "experts" still played the role of "leaders" to help spread the word and guide other users' efforts.

The involvement of slacktivists, influencers, and experts and their complex dynamics call for careful design considerations to enable effective participation of various types of users in user-driven algorithm audits. This includes a) providing affordances for easy, quick, and incidental participation for slacktivists, b) encouraging the engagement of influencers and experts in the auditing process by providing incentivization mechanisms such as bounties as well as feedback about the impact of their engagement, and c) designing communication channels between "leaders" (e.g. influencers and experts) and slacktivists to better guide the auditing effort.

## 7.3 Designing for Participation in User-Driven Algorithm Audits

One of our long-term goals is to design tools specifically intended to help users audit other platforms. Given this context, we discuss some design issues for tools to support user-driven auditing.

*7.3.1* **Designing for Discussion and Deliberation**. Our study showed that users participated in conducting algorithm audits by engaging in threads of conversation which helped them to build on others' hypotheses, ideas, and discussions. However, Table 1 shows that each case had several hundred unique conversations, strongly suggesting that these conversation threads were mostly fragmented and decentralized. That is, one user might tweet out a comment about possible bias in a case along with evidence. The next day, another person might independently do the same, but not realize that another person has found the same problem. Twitter's hashtags are insufficient here since it requires people to use the same hashtags. This fragmentation makes it difficult for a user to see all of the hypotheses put forward, as well as all of the evidence that has been collected.

One implication for design is that any tools for supporting user-driven auditing should consider how to make it easy for people to collect data, see the different hypotheses, the evidence for and against those hypotheses, and related discussion all in a single place.

However, a related challenge is with the large quantity of data. As we saw, there were sometimes thousands of tweets in a case. Aggregating all of this in one place would help people understand how a case unfolded, but make it hard to see, for instance, just the evidence. Furthermore, while some kinds of algorithmic bias are so egregious that one example is sufficient (e.g. computer vision systems labeling Black people as "gorillas"), many other kinds of algorithmic bias may require multiple comparisons before being able to make a conclusion. For example, it would be hard to draw a conclusion from a single or even a few examples of Twitter photo cropping or Apple Card case, but with multiple examples a clear pattern emerges. As such, another implication for design is that tools should make it easy to see multiple instances of evidence at the same time, to make it easier to do comparisons and see potential patterns.

*7.3.2* **Tackling the Access/Visibility Barriers**. One challenge that can slow or hinder a user-driven auditing is lack of sufficient access/visibility to the algorithmic system and difficulty of testing. Access to a service is a known issue for auditing in general [17], but is especially acute for user-driven audits where "not all algorithmic behaviors are equally visible" [23], and therefore testable for users. For example, the Twitter Image Cropping, ImageNet Roulette, and Portrait AI cases were easy for participants to test. The software was widely available, and bias was easy for people to assess. However, Apple Card was limited in that one would have to be married and also apply for the credit card, which could have impact on one's credit rating. For other systems, the access of an impacted user to a service might not be even feasible; for example, in many high-stakes algorithmic decision making systems (such as housing assessment tools, child welfare prediction systems, and predictive policing), the users who are mostly impacted by these systems such as marginalized communities have no access to (and sometimes no awareness of the existence of) the algorithmic process, let alone to the ability of testing and auditing the system [23].

This spectrum of access barriers calls for design considerations and processes that provide users with more visibility into the presence and operation of opaque algorithmic systems. For example, Eslami et al. developed a tool that provided Facebook users with visibility into which of their friends' posts were filtered by Facebook's feed curation algorithm [26], making it easier for users to evaluate and test the algorithm. In another example, to help identify a broader range of issues, Twitter made their image cropping code open source via a bias bounty challenge [15]. This resulted in the detection several other issues in addition to the already detected racial bias.

This added visibility, coming from the platform itself or from third-party entities, can aid users in understanding the mechanisms of an algorithmic system better, and therefore, test it more effectively. In addition, such visibility and access can capture people's attention in a way that motivates them to contribute and share with others. In particular, past work [66] has shown that emotionally charged Twitter messages are more likely to be retweeted, and harmful biases that lead to outrage and anger may be more likely to spread and have more people join the audit.

A potential complementary approach to testing is to gain access to underlying data so that participants can inspect large numbers of results. However, failure to set up the appropriate policy structures could render this approach ineffective. For example, with the Apple Card case, the New York State Department of Financial Services ultimately concluded that there was no observed bias with Goldman Sachs' algorithm. In response, our team filed a Freedom of Information request, from which we learned that the state lost the investigation files and that the records they retained did not include any data that indicated the gender of applicants (see Appendix for response letter). New York State also notified Goldman Sachs' of our open records request. In response, the investment bank asked to continue withholding the investigative records from public inspection.[15] This unsuccessful request shows the need for establishing the right policy and regulation structures for algorithm auditing and accountability that we discuss in the next section.

### 7.3.3 Developing Policies and Regulations for Algorithm Auditing & Accountability.
As the previous section illustrated, increasing visibility/access to algorithmic systems is not always an easy process. There is usually opposition from organizations to providing visibility/transparency into algorithmic systems, in addition to the inherent challenges of providing visibility/transparency to users [4].

Regulatory agencies are one possible point of leverage here. That is, requiring organizations running these algorithmic services to be more accountable and more open may facilitate participation in user-driven algorithm audits and lead to more opportunities for users to detect and report bias in scale. The development of policies by regulatory agencies should also include the stages after a bias is detected. If organizations aim to build legitimacy and empower users, they need to provide visibility into the processes they follow to improve a reported issue. Twitter updating its users about the racial bias of its image cropping algorithm is a good example [15].

A final issue for user-driven audits is the response by organizations. In particular, Twitter's response to concerns about image cropping was to remove the feature. Other companies have had similar responses when people found harmful biases. For example, Microsoft, Amazon, and IBM all halted access to their face recognition services after Buolamwini and Gebru's research showed that these services had more errors for Black people and especially Black women [10]. In the context of assistive technologies, Bennett et al describe how people with visual disabilities saw many benefits but also expressed concerns over possible bias, errors, and discrimination from computer vision systems that could describe people's race, gender, and disabilities in pictures [5]. However, despite the possible benefits, the company building this computer vision system opted not to deploy it, likely due to concerns about negative publicity. On the one hand, in each of these examples the harmful bias was removed. On the other hand, access to potentially useful services was blocked, which may be problematic for situations where certain kinds of errors are tolerable or can be mitigated. In summary, what a company should do in response to a harmful bias can be complicated, and worth much more debate among researchers, ethicists, policy makers, and practitioners.

## 8 CONCLUSION

In this paper, we offered a comparative study of four highly visible user-driven auditing cases on Twitter. Our analysis has shown that they shared similar patterns of user participation, including a large burst of activities followed by a long tail of low activity; and most users contributed only one tweet. We also found a clear division between "content contributors" and "content broadcasters," suggesting both roles are essential for conducting user-driven audits. Finally, we observed that all four cases contained similar types of user-generated content, including hypothesizing, evidence collection, amplification, contextualization, and escalation. Our findings revealed a number of unique engagement patterns of user-driven audits on Twitter and shed light on how to carefully scaffold these spontaneous and fluid collective actions in the future. With respect to future work, one direction is to build tools to facilitate these kinds of user-driven audits. We saw a division of labor with specific kinds of roles. How can this division of labor be done more effectively? How can we incentivize enough people and a diverse enough crowd to participate? How can we support these participants in collecting data, hypothesizing, and analyzing the results? Some of these questions might need other research methods to answer, such as in-depth interviews. We leave them for future work.

## REFERENCES
[1] [n.d.]. *"Apple Card launches today for all US customers"*. https://www.apple.com/newsroom/2019/08/apple-card-launches-today-for-all-us-customers/
[2] 2020. Twitter investigates racial bias in image previews. *BBC News* (Sep 2020). https://www.bbc.com/news/technology-54234822
[3] Parag Agrawal and Dantley Davis. 2020. Transparency around image cropping and changes to come. *Twitter* (2020). https://blog.twitter.com/en_us/topics/product/2020/transparency-image-cropping.html
[4] Mike Ananny and Kate Crawford. 2018. Seeing without knowing: Limitations of the transparency ideal and its application to algorithmic accountability. *new media & society* 20, 3 (2018), 973–989.
[5] Cynthia L Bennett, Cole Gleason, Morgan Klaus Scheuerman, Jeffrey P Bigham, Anhong Guo, and Alexandra To. 2021. "It's Complicated": Negotiating Accessibility and (Mis) Representation in Image Descriptions of Race, Gender, and Disability. In *Proceedings of the 2021 CHI Conference on Human Factors in Computing Systems*. 1–19.
[6] Reuben Binns, Max Van Kleek, Michael Veale, Ulrik Lyngs, Jun Zhao, and Nigel Shadbolt. 2018. 'It's Reducing a Human Being to a Percentage'; Perceptions of Justice in Algorithmic Decisions. *CoRR* abs/1801.10408 (2018). arXiv:1801.10408 http://arxiv.org/abs/1801.10408
[7] Tolga Bolukbasi, Kai-Wei Chang, James Y Zou, Venkatesh Saligrama, and Adam T Kalai. 2016. Man is to computer programmer as woman is to homemaker? debiasing word embeddings. *Advances in neural information processing systems* 29 (2016), 4349–4357.
[8] Virginia Braun and Victoria Clarke. 2006. Using thematic analysis in psychology. *Qualitative Research in Psychology* 3, 2 (2006), 77–101.
[9] Anna Brown, Alexandra Chouldechova, Emily Putnam-Hornstein, Andrew Tobin, and Rhema Vaithianathan. 2019. Toward algorithmic accountability in public services: A qualitative study of affected community perspectives on algorithmic decision-making in child welfare services. In *Proceedings of the 2019 CHI Conference on Human Factors in Computing Systems*. 1–12.
[10] Joy Buolamwini and Timnit Gebru. 2018. Gender Shades: Intersectional Accuracy Disparities in Commercial Gender Classification. In *FAT*.
[11] Ángel Alexander Cabrera, Abraham J Druck, Jason I Hong, and Adam Perer. 2021. Discovering and validating ai errors with crowdsourced failure reports. *Proceedings of the ACM on Human-Computer Interaction* 5, CSCW2 (2021), 1–22.

---

[15]The investment bank claimed the authors, university researchers, were a grave competitive threat to their bottom line, if they had access to a subset of the data their algorithm used to make credit determinations.

[12] Nolan L Cabrera, Cheryl E Matias, and Roberto Montoya. 2017. Activism or slacktivism? The potential and pitfalls of social media in contemporary student activism. *Journal of Diversity in Higher Education* 10, 4 (2017), 400.
[13] Zhilong Chen, Hancheng Cao, Fengli Xu, Mengjie Cheng, Tao Wang, and Yong Li. 2020. Understanding the Role of Intermediaries in Online Social E-commerce: An Exploratory Study of Beidian. *Proceedings of the ACM on Human-Computer Interaction* 4, CSCW2 (2020), 1–24.
[14] Myoung-Gi Chon and Hyojung Park. 2020. Social media activism in the digital age: Testing an integrative model of activism on contentious issues. *Journalism & Mass Communication Quarterly* 97, 1 (2020), 72–97.
[15] Rumman Chowdhury and Jutta Williams. 2021. Introducing Twitter's first algorithmic bias bounty challenge. *Twitter* (2021). https://blog.twitter.com/engineering/en_us/topics/insights/2021/algorithmic-bias-bounty-challenge
[16] Seth Cooper, Firas Khatib, Adrien Treuille, Janos Barbero, Jeehyung Lee, Michael Beenen, Andrew Leaver-Fay, David Baker, Zoran Popović, et al. 2010. Predicting protein structures with a multiplayer online game. *Nature* 466, 7307 (2010), 756–760.
[17] Sasha Costanza-Chock, Inioluwa Deborah Raji, and Joy Buolamwini. 2022. Who Audits the Auditors? Recommendations from a Field Scan of the Algorithmic Auditing Ecosystem. In *2022 ACM Conference on Fairness, Accountability, and Transparency* (Seoul, Republic of Korea) *(FAccT '22)*. Association for Computing Machinery, New York, NY, USA, 1571–1583. https://doi.org/10.1145/3531146.3533213
[18] Justin Cranshaw and Aniket Kittur. 2011. The polymath project: lessons from a successful online collaboration in mathematics. In *Proceedings of the SIGCHI conference on human factors in computing systems*. 1865–1874.
[19] Kate Crawford and Trevor Paglen. 2021. Excavating AI: The politics of images in machine learning training sets. *AI & SOCIETY* (2021), 1–12.
[20] Jia Deng, Wei Dong, Richard Socher, Li-Jia Li, Kai Li, and Li Fei-Fei. 2009. ImageNet: A large-scale hierarchical image database. In *2009 IEEE conference on computer vision and pattern recognition*. IEEE, 248–255.
[21] Michael A. DeVito, Jeremy Birnholtz, Jeffery T. Hancock, Megan French, and Sunny Liu. 2018. How People Form Folk Theories of Social Media Feeds and What It Means for How We Study Self-Presentation. In *Proceedings of the 2018 CHI Conference on Human Factors in Computing Systems* (Montreal QC, Canada) *(CHI '18)*. Association for Computing Machinery, New York, NY, USA, 1–12. https://doi.org/10.1145/3173574.3173694
[22] Michael A DeVito, Darren Gergle, and Jeremy Birnholtz. 2017. "Algorithms ruin everything" # RIPTwitter, Folk Theories, and Resistance to Algorithmic Change in Social Media. In *Proceedings of the 2017 CHI conference on human factors in computing systems*. 3163–3174.
[23] Alicia DeVos, Aditi Dhabalia, Hong Shen, Kenneth Holstein, and Motahhare Eslami. 2022. Toward User-Driven Algorithm Auditing: Investigating users' strategies for uncovering harmful algorithmic behavior. In *CHI Conference on Human Factors in Computing Systems*. 1–19.
[24] John Downey and Natalie Fenton. 2003. New media, counter publicity and the public sphere. *New media & society* 5, 2 (2003), 185–202.
[25] Motahhare Eslami, Karrie Karahalios, Christian Sandvig, Kristen Vaccaro, Aimee Rickman, Kevin Hamilton, and Alex Kirlik. 2016. First I" like" it, then I hide it: Folk Theories of Social Feeds. In *Proceedings of the 2016 cHI conference on human factors in computing systems*. 2371–2382.
[26] Motahhare Eslami, Aimee Rickman, Kristen Vaccaro, Amirhossein Aleyasen, Andy Vuong, Karrie Karahalios, Kevin Hamilton, and Christian Sandvig. 2015. " I always assumed that I wasn't really that close to [her]" Reasoning about Invisible Algorithms in News Feeds. In *Proceedings of the 33rd annual ACM conference on human factors in computing systems*. 153–162.
[27] Motahhare Eslami, Kristen Vaccaro, Karrie Karahalios, and Kevin Hamilton. 2017. "Be careful; things can be worse than they appear": Understanding Biased Algorithms and Users' Behavior around Them in Rating Platforms. In *Proceedings of the International AAAI Conference on Web and Social Media*, Vol. 11.
[28] Motahhare Eslami, Kristen Vaccaro, Min Kyung Lee, Amit Elazari Bar On, Eric Gilbert, and Karrie Karahalios. 2019. User attitudes towards algorithmic opacity and transparency in online reviewing platforms. In *Proceedings of the 2019 CHI Conference on Human Factors in Computing Systems*. 1–14.
[29] Casey Fiesler and Nicholas Proferes. 2018. "Participant" perceptions of Twitter Research Ethics. *Social Media + Society* 4, 1 (2018), 205630511876336. https://doi.org/10.1177/2056305118763366
[30] Kristie Fisher, Scott Counts, and Aniket Kittur. 2012. Distributed sensemaking: Improving sensemaking by leveraging the efforts of previous users. In *Proceedings of the 2012 CHI Conference on Human Factors in Computing Systems*. 247–256.
[31] Nancy Fraser. 2021. Rethinking the public sphere: A contribution to the critique of actually existing democracy. In *Public Space Reader*. Routledge, 34–41.
[32] Orit Gat. 2019. "How AI Selfie App ImageNet Roulette Took the Internet by Storm". (2019). https://www.frieze.com/article/how-ai-selfie-app-imagenet-roulette-took-internet-storm
[33] R Stuart Geiger and Aaron Halfaker. 2013. Using edit sessions to measure participation in Wikipedia. In *Proceedings of the 2013 Conference on Computer Supported Cooperative Work*. 861–870.
[34] Susan A Gelman and Cristine H Legare. 2011. Concepts and folk theories. *Annual review of anthropology* 40 (2011), 379–398.
[35] Carol Gilligan. 1993. *In a different voice: Psychological theory and women's development*. Harvard University Press.
[36] Nina Grgic-Hlaca, Elissa M. Redmiles, Krishna P. Gummadi, and Adrian Weller. 2018. Human Perceptions of Fairness in Algorithmic Decision Making: A Case Study of Criminal Risk Prediction. (2018), 903–912. https://doi.org/10.1145/3178876.3186138
[37] Gabriel Grill and Nazanin Andalibi. 2022. Attitudes and Folk Theories of Data Subjects on Transparency and Accuracy in Emotion Recognition. *Proc. ACM Hum.-Comput. Interact.* 6, CSCW1, Article 78 (apr 2022), 35 pages. https://doi.org/10.1145/3512925
[38] Aaron Halfaker and R Stuart Geiger. 2020. Ores: Lowering barriers with participatory machine learning in wikipedia. *Proceedings of the ACM on Human-Computer Interaction* 4, CSCW2 (2020), 1–37.
[39] Aaron Halfaker, Aniket Kittur, and John Riedl. 2011. Don't bite the newbies: How reverts affect the quantity and quality of Wikipedia work. In *Proceedings of the 7th International Symposium on Wikis and Open Collaboration*. 163–172.
[40] Aniko Hannak, Gary Soeller, David Lazer, Alan Mislove, and Christo Wilson. 2014. Measuring price discrimination and steering on e-commerce web sites. In *Proceedings of the 2014 Conference on Internet Measurement Conference*. 305–318.
[41] Alex Hern. 2020. Twitter apologises for 'racist' image-cropping algorithm. *The Guardian* (Sept. 2020). https://www.theguardian.com/technology/2020/sep/21/twitter-apologises-for-racist-image-cropping-algorithm
[42] Khari Johnson. 2020. Apparent racial bias found in Twitter photo algorithm. *VentureBeat* (Sep 2020). https://venturebeat.com/2020/09/20/apparent-racial-bias-found-in-twitter-photo-algorithm/
[43] Nadia Karizat, Dan Delmonaco, Motahhare Eslami, and Nazanin Andalibi. 2021. Algorithmic Folk Theories and Identity: How TikTok Users Co-Produce Knowledge of Identity and Engage in Algorithmic Resistance. *Proc. ACM Hum.-Comput. Interact.* 5, CSCW2, Article 305 (oct 2021), 44 pages. https://doi.org/10.1145/3476046
[44] Sara Kingsley, Proteeti Sinha, Clara Wang, Motahhare Eslami, and Jason I. Hong. 2022. "Give Everybody [..] a Little Bit More Equity": Content Creator Perspectives and Responses to the Algorithmic Demonetization of Content Associated with Disadvantaged Groups. *Proc. ACM Hum.-Comput. Interact.* 6, CSCW2, Article 424 (nov 2022), 37 pages. https://doi.org/10.1145/3555149
[45] Sara Kingsley, Clara Wang, Alex Mikhalenko, Proteeti Sinha, and Chinmay Kulkarni. 2020. Auditing Digital Platforms for Discrimination in Economic Opportunity Advertising. https://doi.org/10.48550/ARXIV.2008.09656
[46] Shamika Klassen. 2022. Black Twitter is Gold: Why This Online Community is Worthy of Study and How to Do so Respectfully. *Interactions* 29, 1 (jan 2022), 96–98. https://doi.org/10.1145/3505681
[47] Will Knight. 2019. "The Apple Card Didn't 'See' Gender—and That's the Problem". (2019). https://www.wired.com/story/the-apple-card-didnt-see-genderand-thats-the-problem/
[48] Irene Font Peradejordi Kyra Yee. 2021. "Sharing learnings from the first algorithmic bias bounty challenge". *Twitter* (2021). https://blog.twitter.com/engineering/en_us/topics/insights/2021/learnings-from-the-first-algorithmic-bias-bounty-challenge
[49] Michelle S. Lam, Mitchell L. Gordon, Danaë Metaxa, Jeffrey T. Hancock, James A. Landay, and Michael S. Bernstein. 2022. End-User Audits: A System Empowering Communities to Lead Large-Scale Investigations of Harmful Algorithmic Behavior. *Proc. ACM Hum.-Comput. Interact.* 6, CSCW2, Article 512 (nov 2022), 34 pages. https://doi.org/10.1145/3555625
[50] J Richard Landis and Gary G Koch. 1977. The measurement of observer agreement for categorical data. *biometrics* (1977), 159–174.
[51] Yu-Hao Lee and Gary Hsieh. 2013. Does slacktivism hurt activism? The effects of moral balancing and consistency in online activism. In *Proceedings of the SIGCHI conference on human factors in computing systems*. 811–820.
[52] Danaë Metaxa, Joon Sung Park, Ronald E Robertson, Karrie Karahalios, Christo Wilson, Jeff Hancock, Christian Sandvig, et al. 2021. Auditing algorithms: Understanding algorithmic systems from the outside in. *Foundations and Trends® in Human–Computer Interaction* 14, 4 (2021), 272–344.
[53] Carolina Milanesi. 2021. "The Apple Card's new feature tackles one of credit's biggest problems". (2021). https://www.fastcompany.com/90627796/apple-card-apple-card-family-share-card-children
[54] Safiya Umoja Noble. 2018. Algorithms of Oppression: How Search Engines Reinforce Racism. (2018). https://www.jstor.org/stable/j.ctt1pwt9w5
[55] New York State Department of Financial Services. 2021. "NYSDFS Report on Apple Card Investigation". (2021). https://www.dfs.ny.gov/system/files/documents/2021/03/rpt_202103_apple_card_investigation.pdf
[56] Parmy Olson. 2018. The algorithm that helped google translate become sexist. *Forbes* (Feb. 2018). https://www.forbes.com/sites/parmyolson/2018/02/15/the-algorithm-that-helped-google-translate-become-sexist/
[57] Charles C. Ragin and Howard S. Becker. 1992. *What is a case?: Exploring the foundations of social inquiry*. Cambridge University Press.

[58] Inioluwa Deborah Raji and Joy Buolamwini. 2019. Actionable Auditing: Investigating the Impact of Publicly Naming Biased Performance Results of Commercial AI Products. In *Proceedings of the 2019 AAAI/ACM Conference on AI, Ethics, and Society* (Honolulu, HI, USA) *(AIES '19)*. Association for Computing Machinery, New York, NY, USA, 429–435. https://doi.org/10.1145/3306618.3314244

[59] Brandon Reynante, Steven P Dow, and Narges Mahyar. 2021. A framework for open civic design: Integrating public participation, crowdsourcing, and design thinking. *Digital Government: Research and Practice* 2, 4 (2021), 1–22.

[60] Ronald E. Robertson, Shan Jiang, Kenneth Joseph, Lisa Friedland, David Lazer, and Christo Wilson. 2018. Auditing Partisan Audience Bias within Google Search. *Proc. ACM Hum.-Comput. Interact.* 2, CSCW, Article 148 (Nov. 2018), 22 pages. https://doi.org/10.1145/3274417

[61] Dana Rotman, Sarah Vieweg, Sarita Yardi, Ed Chi, Jenny Preece, Ben Shneiderman, Peter Pirolli, and Tom Glaisyer. 2011. From slacktivism to activism: participatory culture in the age of social media. In *CHI'11 Extended Abstracts on Human Factors in Computing Systems*. 819–822.

[62] Christina Ruiz. 2019. Leading online database to remove 600,000 images after art project reveals its racist bias. *The Art Newspaper* (2019), 23.

[63] Christian Sandvig, Kevin Hamilton, Karrie Karahalios, and Cedric Langbort. 2014. Auditing algorithms: Research methods for detecting discrimination on internet platforms. *Data and discrimination: converting critical concerns into productive inquiry* 22 (2014), 4349–4357.

[64] James C Scott. 1985. *Weapons of the weak: Everyday forms of peasant resistance.* Yale University Press.

[65] Hong Shen, Alicia DeVos, Motahhare Eslami, and Kenneth Holstein. 2021. Everyday Algorithm Auditing: Understanding the Power of Everyday Users in Surfacing Harmful Algorithmic Behaviors. *Proc. ACM Hum.-Comput. Interact.* 5, CSCW2, Article 433 (oct 2021), 29 pages. https://doi.org/10.1145/3479577

[66] Stefan Stieglitz and Linh Dang-Xuan. 2013. Emotions and information diffusion in social media—sentiment of microblogs and sharing behavior. *Journal of management information systems* 29, 4 (2013), 217–248.

[67] Latanya Sweeney. 2013. Discrimination in online ad delivery. *Commun. ACM* 56, 5 (2013), 44–54.

[68] Niels van Berkel, Jorge Goncalves, Danula Hettiachchi, Senuri Wijenayake, Ryan M. Kelly, and Vassilis Kostakos. 2019. Crowdsourcing Perceptions of Fair Predictors for Machine Learning: A Recidivism Case Study. *Proc. ACM Hum.-Comput. Interact.* 3, CSCW, Article 28 (Nov. 2019), 21 pages. https://doi.org/10.1145/3359130

[69] Neil Vigdor. 2019. Apple Card Investigated After Gender Discrimination Complaints. (2019). https://www.nytimes.com/2019/11/10/business/Apple-credit-card-investigation.html

[70] Luis Von Ahn and Laura Dabbish. 2004. Labeling images with a computer game. In *Proceedings of the SIGCHI conference on Human factors in computing systems*. 319–326.

[71] Karl E Weick. 1995. *Sensemaking in organizations.* Vol. 3. Sage.

[72] Allison Woodruff, Sarah E. Fox, Steven Rousso-Schindler, and Jeffrey Warshaw. 2018. A Qualitative Exploration of Perceptions of Algorithmic Fairness. (2018), 1–14. https://doi.org/10.1145/3173574.3174230

[73] Yuxi Wu, W Keith Edwards, and Sauvik Das. 2022. "A Reasonable Thing to Ask For": Towards a Unified Voice in Privacy Collective Action. In *CHI Conference on Human Factors in Computing Systems*. 1–17.

[74] Kyra Yee, Uthaipon Tantipongpipat, and Shubhanshu Mishra. 2021. Image Cropping on Twitter: Fairness Metrics, Their Limitations, and the Importance of Representation, Design, and Agency. *Proc. ACM Hum.-Comput. Interact.* 5, CSCW2, Article 450 (oct 2021), 24 pages. https://doi.org/10.1145/3479594

[75] Cat Zakrzewski. 2021. The Technology 202: Democrats introduce legislation prohibiting algorithmic discrimination. *The Washington Post* (2021).

# A   FREEDOM OF INFORMATION LAW (FOIL) LETTER

This letter is from New York State Department of Financial Services, in response to our request for data about the Apple Card case. The response letter is addressed to Goldman Sachs' lawyer, despite the fact the letter was requested by one of the co-authors of this submission. Since the letter is a public record produced by New York state government, we have included all name information in the original letter instead of redacting it (during peer-review the co-author's name who made the request was redacted).

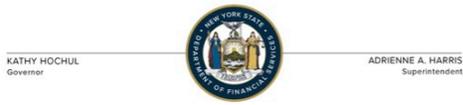

KATHY HOCHUL                    ADRIENNE A. HARRIS
Governor                        Superintendent

**VIA EMAIL**
(MDenerstein@gibsondunn.com)

June 2, 2022

Mylan L. Denerstein
Gibson Dunn & Crutcher LLP
200 Park Avenue
New York, NY 10166-0193

Re:   Freedom of Information Law ("FOIL") Request - Tracking No. 2022-089785

Dear Ms. Denerstein:

Pursuant to N.Y. Public Officers Law § 89(5)(b)(3), this correspondence represents the determination of the New York State Department of Financial Services ("Department") of the request by your client, Goldman Sachs Bank USA ("Goldman"), that records it submitted to the Department continue to be excepted from FOIL disclosure under N.Y. Public Officers Law § 87(2)(d). For the reasons set forth below, your client's request is granted.

**I.   Background**

The Department received the following FOIL request on January 25, 2022:

> Dear NYSDFS, We request a copy of the data in this report: https://www.dfs.ny.gov/system/files/documents/2021/03/rpt_2021 03_apple_card_investigation.pdf Specifically, we request the records for the 400,000 applicants, including these variables: gender, race, age, salary or income, credit score, credit limit and if possible zip code.

The Department construes this request to seek records containing data of applicants submitted to and reviewed by the Department during its investigation of Goldman Sachs' underwriting of Apple-branded credit cards, which includes the gender, race, age, salary, income, credit score, credit limit and zip codes of applicants for such card. After a diligent search of the Department's records, a record containing all such data could not be located.

THE NEW YORK STATE DEPARTMENT OF FINANCIAL SERVICES
EQUITABLE • INNOVATIVE • COLLABORATIVE • TRANSPARENT

---

Mylan L. Denerstein
June 2, 2022
Page 2 of 4

**II.   Trade Secret Determination**

Notwithstanding that determination, the Department has located a record that contains some of the demographics mentioned in the request, which is a spreadsheet of applicant data that includes a) application identification numbers, b) applicant names, c) applicant dates of birth, d) FICO scores, e) zip codes, and f) application decisions ("Applicant Data"). When Goldman submitted the Applicant Data to the Department, Goldman requested that it be excepted from public disclosure pursuant to Public Officers Law § 87(2)(d). That statute provides an exception from disclosure for records or portions thereof that "are trade secrets or are submitted to an agency by a commercial enterprise or derived from information obtained from a commercial enterprise and which if disclosed would cause substantial injury to the competitive position of the subject enterprise."

Thereafter, the Department requested that Goldman submit a Statement of Necessity ("Statement") pursuant to Public Officers Law § 89(5). Goldman submitted its Statement requesting that the Department continue to except the Applicant Data from public disclosure. In the Statement, Goldman asserts that the Applicant Data is protected under Public Officers Law § 87(2)(d), since disclosure would cause substantial injury to Goldman's competitive position. Specifically, Goldman asserts that the Applicant Data contains its proprietary information, which includes its confidential credit underwriting policies and procedures, as well as voluminous records reflecting personal and financial data of applicants that Goldman Sachs deemed relevant to its confidential credit underwriting analyses. Accordingly, Goldman asserts that disclosure of its Applicant Data would cause it to sustain substantial competitive injury, since it would allow competitors to offer more favorable terms, conditions, rates, and benefits to customers without undertaking the significant time, effort and expense expended by Goldman to develop and test its underwriting models and formulas. Consequently, disclosing the Applicant Data would allow competitors to outcompete Goldman, thereby causing Goldman to sustain substantial competitive injury.

In Encore Coll. Bookstores, Inc. v. Auxiliary Serv. Corp., 87 N.Y.2d 410 (1995), the New York Court of Appeals provided the standard for determining whether records responsive to a FOIL request should be withheld from disclosure due to competitive injury concerns. In Encore, the Court held that determining whether an entity suffers competitive injury from public disclosure of its records within the meaning of Public Officers Law § 87(2)(d), "turns on the commercial value of the requested information to competitors and the cost of acquiring it through other means." Id. at 420. The Court's holding recognized that the trade secret/competitive injury exemption balances the need for openness in government versus the need "to protect businesses from the deleterious consequences of disclosing confidential commercial information, so as to further [New York] State's economic development efforts and attract business to New York." Id. Whether disclosing records, in whole or in part, would cause substantial competitive injury to an entity depends on, among other things, the nature of the information involved and the area of commerce in which the entity does business. See Committee on Open Government, Advisory Opinion 10664 (March 10, 1998).

---

Mylan L. Denerstein
June 2, 2022
Page 3 of 4

Applying Encore, the Department agrees that disclosure of the Applicant Data would cause Goldman to suffer substantial competitive injury. A competitor who submits a FOIL request for the Applicant Data would certainly receive a windfall, since the competitor would have paid only a minimal fee, if any, to obtain proprietary information that Goldman has expended substantial resources to develop. Additionally, disclosing the Applicant Data would create an unlevel playing field, as Goldman does not possess the proprietary underwriting policies and procedures or applicant demographic information of its competitors. Finally, releasing the Applicant Data would undermine Goldman's business interests and result in competitive harm, since Goldman's customers rely on its ability to preserve the confidentiality of its business transactions and customer relationships and new and existing customers alike would move their business to banks that are able to preserve confidentiality. See N.Y. State Elec. & Gas Corp. v. NYS Energy Planning Board, 221 A.D.2d 121, 125 (3rd Dep't 1996) (holding that respondent would be unable to fairly compete for customers in the future if the respondent could not guarantee that customer confidentiality would not be violated). Therefore, disclosing the Applicant Data would unfairly benefit Goldman's competitors, thereby causing Goldman to suffer substantial injury to its competitive position.

Based upon Goldman's Statement, and pursuant to Public Officers Law § 87(2)(d), the Department determines that the Applicant Data will not be disclosed, since disclosure would cause substantial injury to Goldman's competitive position.

Pursuant to Public Officers Law § 89(5)(c), the requester may appeal the **Trade Secret Determination within seven business days** by sending an email to FOIL.Appeals@dfs.ny.gov.

**III.   Final Determination**

Furthermore, the Department is withholding the Applicant Data from disclosure pursuant to Public Officers Law §§ 87(2)(a) and (b). Specifically, Public Officers Law § 87(2)(a) exempts from disclosure records exempted by a state or federal statute. The applicable statutory provision here is N.Y. Banking Law ("Banking Law") § 36(10). Banking Law § 36(10) states, in pertinent part, that reports of examinations and investigations and correspondence and memoranda concerning or arising out of such examinations and investigations are confidential and shall not be made public. The statute fosters open communication between the Department and its regulated institutions, a necessity for effective regulation of financial institutions, by ensuring that records transmitted to the Department in connection with its supervision of a financial institution are protected from disclosure. The Applicant Data were transmitted to the Department in connection with the Department's investigation of Goldman Sachs' underwriting of Apple-branded credit cards and are, therefore, required to be kept confidential and exempt from disclosure under Banking Law § 36(10). Release of such sensitive records may have a chilling effect on the willingness of regulated entities to share information and cooperate with supervisors to resolve issues, thus, confidentiality is critical to the Department's ability to perform its regulatory mandate and purpose. Accordingly, the exemption set forth in Public Officers Law § 87(2)(a) and Banking Law § 36(10) covers the Applicant Data.

---

Mylan L. Denerstein
June 2, 2022
Page 4 of 4

Public Officers Law § 87(2)(b) (the "Personal Privacy Exemption") prohibits the disclosure of responsive records or portions thereof that "if disclosed would constitute an unwarranted invasion of personal privacy under the provisions of [Public Officers Law § 89(2)]." Determining whether disclosure would constitute an unwarranted invasion of personal privacy requires balancing the competing interests of public access and individual privacy. See Dobranski v. Houper, 154 A.D.2d 736 (3d Dep't 1989).

Here, the Department is withholding the Applicant Data for containing information such as application identification numbers and dates of births, since the interest in keeping that information private outweighs the competing interest of providing public access to that information. Disclosing this information about private citizens would not serve any governmental purpose consistent with the intent of FOIL. See Goyer v. New York State Dep't of Envtl. Conservation, 12 Misc.3d 261 (Sup. Ct. N.Y. County 2005).

For the above-mentioned reasons, the Department has denied the FOIL request. Pursuant to Public Officers Law § 89(4), the requester may appeal the **Final Determination within thirty days** by sending an email to FOIL.Appeals@dfs.ny.gov.

Very truly yours,

Norma Roper Webster
Senior Attorney & Record Access Officer

cc:   **VIA EMAIL**

(Shoshanah.Bewlay@dos.ny.gov)
Shoshanah Bewlay, Executive Director
Committee on Open Government
One Commerce Plaza
99 Washington Avenue-Suite 650
Albany, NY 12231

(skingsle@cs.cmu.edu)
Ms. Sara Kingsley
Carnegie Mellon University
5000 Forbes Avenue
Pittsburgh, PA 15213